\documentclass[12pt]{article}
\usepackage{latexsym}
\usepackage{amssymb}
\usepackage[dvips]{graphicx}
\usepackage{epsfig}
\usepackage{rotating}
\usepackage{amsfonts}
\usepackage{amsmath}

\usepackage{hyperref}
\usepackage{pstricks}

\newtheorem{definition}{Definition}[section]

\newtheorem{prop}{Proposition}[section]

\newtheorem{lemma}[prop]{Lemma}

\newcommand{\beqa}{\begin{eqnarray}}
\newcommand{\eeqa}{\end{eqnarray}}

\newcommand{\dd}{{\mathrm d}}
\newcommand{\eqn}[1]{(\ref{#1})}

\newcommand{\bea}{\begin{eqnarray}}
\newcommand{\eea}{\end{eqnarray}}
\def\nn{\nonumber}

\newcommand{\TT}{{\mathbb T^{\gamma}}}
\newcommand{\jj}{{\mathbb P}}

\newcommand{\cG}{{\mathcal G}}

\newcommand{\SU}{{\rm SU}}

\newcommand{\be}{\begin{equation}}
\newcommand{\ee}{\end{equation}}
\newcommand{\beq}{\begin{eqnarray}}
\newcommand{\eeq}{\end{eqnarray}}

\newcommand{\bra}{\langle}
\newcommand{\ket}{\rangle}

\newcommand{\rd}{\mathrm{d}}
\newcommand{\tr}{{\rm Tr}}

\def\lsim{\
  \lower-2.0pt\vbox{\hbox{\rlap{$<$}\lower5.5pt\vbox{\hbox{$\sim$}}}}\ }
\def\gsim{\
  \lower-2.0pt\vbox{\hbox{\rlap{$>$}\lower5.5pt\vbox{\hbox{$\sim$}}}}\ }

\begin{document}

\begin{titlepage}
\begin{flushright}
\end{flushright}

\vspace{20pt}

\begin{center}

{\Large\bf Quantum Corrections in the\\
\vspace{10pt}
Group Field Theory Formulation\\
\vspace{10pt}
of the EPRL/FK Models}
\vspace{20pt}

Thomas Krajewski$^{a,b,(1)}$,
Jacques Magnen$^{c,a,(2)}$\\
Vincent Rivasseau $^{a,(3)}$, Adrian Tanasa$^{d,e, (4)}$\\
Patrizia Vitale$^{a,f, (5)}$

\vspace{15pt}

$^{a}${\sl Laboratoire de Physique Th\'eorique, Universit\'e Paris XI}\\
{\sl CNRS UMR 8627, 91405 Orsay Cedex, France}\\
\vspace{10pt}
$^{b}${\sl on leave, Centre de Physique Th\'eorique, CNRS UMR 6207}\\
{\sl CNRS Luminy, Case 907, 13288 Marseille Cedex 9}\\
\vspace{10pt}
$^{c}${\sl Centre de Physique Th\'eorique, Ecole Polytechnique}\\
{\sl CNRS UMR 7644, 91128 Palaiseau Cedex, France}\\

\vspace{10pt}$^{d}${\sl
LIPN, Institut Galil\'ee, CNRS UMR 7030\\
Univ. Paris Nord, 99 av. Cl\'ement, 93430 Villetaneuse, France}\\

\vspace{10pt}
$^{e}${\sl
Institutul de Fizic\u a \c si Inginerie Nuclear\u a Horia Hulubei,\\
P. O. Box MG-6, 077125 M\u agurele, Rom\^ania}\\

\vspace{10pt}$^{f}${\sl
Dipartimento di Scienze Fisiche, Universit\`a di Napoli Federico II\\
and INFN, Sezione di Napoli, Via Cintia 80126 Napoli, Italy}\\

\vspace{20pt}
E-mail:  $^{(1)}${\em krajew@cpt.univ-mrs.fr} , \quad
$^{(2)}${\em magnen@cpht.polytechnique.fr},
\quad $^{(3)}${\em rivass@th.u-psud.fr },\quad $^{(4)}${\em adrian.tanasa@ens-lyon.org},\quad $^{(5)}${\em vitale@na.infn.it}

\vspace{10pt}

\begin{abstract}
\noindent
We investigate the group field theory formulation of
the EPRL/FK spin foam models. These models
aim at a dynamical, i.e. non-topological formulation of 4D quantum gravity.
We introduce a saddle point method for general group field theory
amplitudes and compare it with existing results, in particular
for a second order correction to the EPRL/FK propagator.
\end{abstract}

\end{center}

\noindent  Pacs numbers:   04.60.-m, 04.60.Pp

\end{titlepage}


\setcounter{footnote}{0}

\section{Introduction}
\label{Intro}
\renewcommand{\theequation}{\thesection.\arabic{equation}}
\setcounter{equation}{0}

Group field theories (GFTs) \cite{boul} are quantum field theories
over group manifolds and can be also viewed as higher rank tensor
field theories \cite{Freidel,oriti} which generalize matrix models.
They provide one of the most promising frameworks for a background
invariant theory of quantum gravity in which one sums both over
topologies and geometries. Indeed, each Feynman graph of a $D$
dimensional GFT can be dually associated with a discrete space-time
via a specific triangulation and gluing rules given by the
covariance and vertices of the theory. The functional integral
formalism defines  a weighted sum over triangulations with each
weight (amplitude) related to a sum over geometries via a spin foam
formalism \cite{rovel} (see \cite{FreiLoua, MNRS} for results on
power counting and non-perturbative resummation of such models).

Spin foams are the Feynman amplitudes of GFT. But GFT in addition specifies the
class of graphs that should be summed, together with their combinatoric factors.
This stems from Wick contractions rules, hence (perturbative) GFT requires to distinguish
the non-quadratic part (interaction) from the quadratic part (propagator)
in the field action.

The simplest group field theories correspond to quantization of the BF models,
hence to topological versions of gravity.
Recently new spin foam rules have been proposed for the quantization of full fledged 4D gravity
\cite{EPR,LS,FreKra,ELPR}. These models stem from an improved
analysis of the Plebanski simplicity constraints.
The corresponding so-called EPRL or FK
models are neither of the BF nor of the Barret-Crane type. They
mix the left and right part of $SO(4)\simeq SU(2)\times SU(2)$ in a new  way which gives a central r\^ole to the Immirzi parameter.
These new theories could be called dynamical since their
propagators, combining two non-commuting
projectors, have non-trivial spectrum.

Preliminary studies of the asymptotic large spin (also called ``ultraspin'') regime
have been performed  for the EPRL/FK amplitudes of
the "self-energy" and the "starfish" graphs (see Figures \ref{2P-fig} and \ref{5P-fig}) \cite{PRS}.
These results are a first step towards a study of renormalizability of such theories.

In \cite{linhom} a linearized approximation has been devised
to investigate the ultraspin limit of BF spin foam amplitudes
(see also \cite{BonSme}). This approximation captures the correct
power counting of some graphs, such as type 1 graphs in the Boulatov model \cite{fgo}, but
it typically overestimates more general graphs.

In this paper we push further the group field theory approach to the
 EPRL/FK models, first introduced in \cite{FreKra}, and perform another step towards the
general investigation of their renormalizability. We use a coherent
state representation of the EPRL/FK propagator, as in \cite{FreKra},
while  other representations are exhibited mainly for comparison
with other approaches. We introduce a general saddle point
approximation,  as in \cite{FreidelConrady}, which reproduces
correctly the approximation \cite{linhom} to the power counting of
BF amplitudes for simply connected graphs and, for non-degenerate
configurations,  the EPRL/FK ``self-energy graph'' power counting of
\cite{PRS}. We discuss also the case of degenerate configurations,
not studied in \cite{PRS}.

The plan is as follows: the next section is devoted to a review of
the BF and EPRL/FK group field models in a field theoretical spirit.
The following section presents the stationary
phase method. Finally we remark that the sign of the self-energy
graph points towards a singularity in the effective propagator of
the EPRL/FK model, which could signal a phase transition. For
completeness we included useful formulas and normalization
conventions in the appendices.

\section{Implementation of GFT}
\renewcommand{\theequation}{\thesection.\arabic{equation}}
\setcounter{equation}{0}

In GFT, the field arguments live on products of Lie groups.
Feynman amplitudes are spin-foams, {\it i.e.} two-complexes with vertices,
stranded lines (also called propagators) and faces (that is, closed circuits of strands).

\subsection{Fields}

Since GFT represents a quantum theory of space-time itself, the usual spin-statistic theorem may not apply.
In this paper we consider only Bosonic statistics. However other choices have been considered \cite{gurau1}.
We also work with an Euclidean signature.

The number of strands in the GFT lines encodes the space-time dimension $D$.
The natural group associated to such a $D$ dimensional GFT is $[SO(D)]^D$, hence a
field $\phi$ is a function on $[SO(D)]^D$. We don't assume any symmetry under
permutations of the arguments.

\subsection{Vertices}

In the spin-foam literature the term ``vertex'' usually refers to the vertex {\it together with the square roots
of its dressing propagators}. This terminology
is not the standard one in quantum field theory. Further confusion often stems from the fact that in
$BF$ theory the propagator is a projector hence is equal to its square, and also to its square root!

To clarify the situation, let us return to ordinary field theory. In that case also the definition of the vertex
could be considered ambiguous
since one can dress it with a more or less arbitrary fraction of the propagator.  What fixes this ambiguity is
the usual requirement that vertices in field theory should obey a certain {\it locality} property in direct space.
This allows to distinguish them from their dressing (half)-propagators, which are non-local operators.

Since GFT is non-local on the group
we cannot transpose directly this rule. To properly distinguish the vertex from
the propagator we propose to use an extended notion of locality adapted to the GFT case, which we call {\it simpliciality} \cite{Tana1}.

For consistency reasons every vertex in GFT is required to have
a total degree in the fields ensuring parity of the number of strands.
In odd dimensions this restricts the degree of the vertex to be even. Hence we propose the following definition:

\begin{definition}
A vertex joining $2p$ strands is called simplicial if it has for kernel in direct group space a product of
$p$ delta functions matching strand arguments, so that each delta function joins two
strands in two different half-lines.
\end{definition}

The usual
vertex for $D-$dimensional GFT is a $\phi^{D+1}$   simplicial vertex
in which the faces are glued in the pattern of a $D$-dimensional simplex.
For instance the ordinary Boulatov vertex in 3 dimensions is simplicial (with $p=6$) as it writes
\beqa
S_{{\mathrm{int}}}[\phi]=\frac{\lambda}{4}\int \left(\prod_{i=1}^{12} \dd g_i\right)\phi(g_1,g_2,g_3) \phi(g_4,g_5,g_6)\phi(g_7,g_8,g_9)\phi(g_{10},g_{11},g_{12})
K(g_1, .. g_{12}),   \label{vertexdirect}
\eeqa
with a kernel
\be  K(g_1, .. g_{12}) = \delta(g_3g_4^{-1})\delta(g_2g_8^{-1})\delta(g_6g_7^{-1})\delta(g_9g_{10}^{-1})\delta(g_5g_{11}^{-1})\delta(g_1g_{12}^{-1})
\ee
satisfying to our definition.
But remark that the ``pillow term'' \cite{FreiLoua}
\beqa
S_{{\mathrm {int}}}^{{\mathrm {pillow}}}[\phi]=\frac{\lambda}{4}\int \left(\prod_{i=1}^{6} \dd g_i\right)\phi(g_1,g_2,g_3) \phi(g_3,g_4,g_5)\phi(g_5,g_4,g_6)\phi(g_6,g_2,g_1).
\eeqa
 is also simplicial in $D=3$.  Also in any dimension $D$
 there are infinitely many higher than order $D+1$ simplicial vertices according to our definition.

\begin{figure}
\centerline{\epsfig{figure=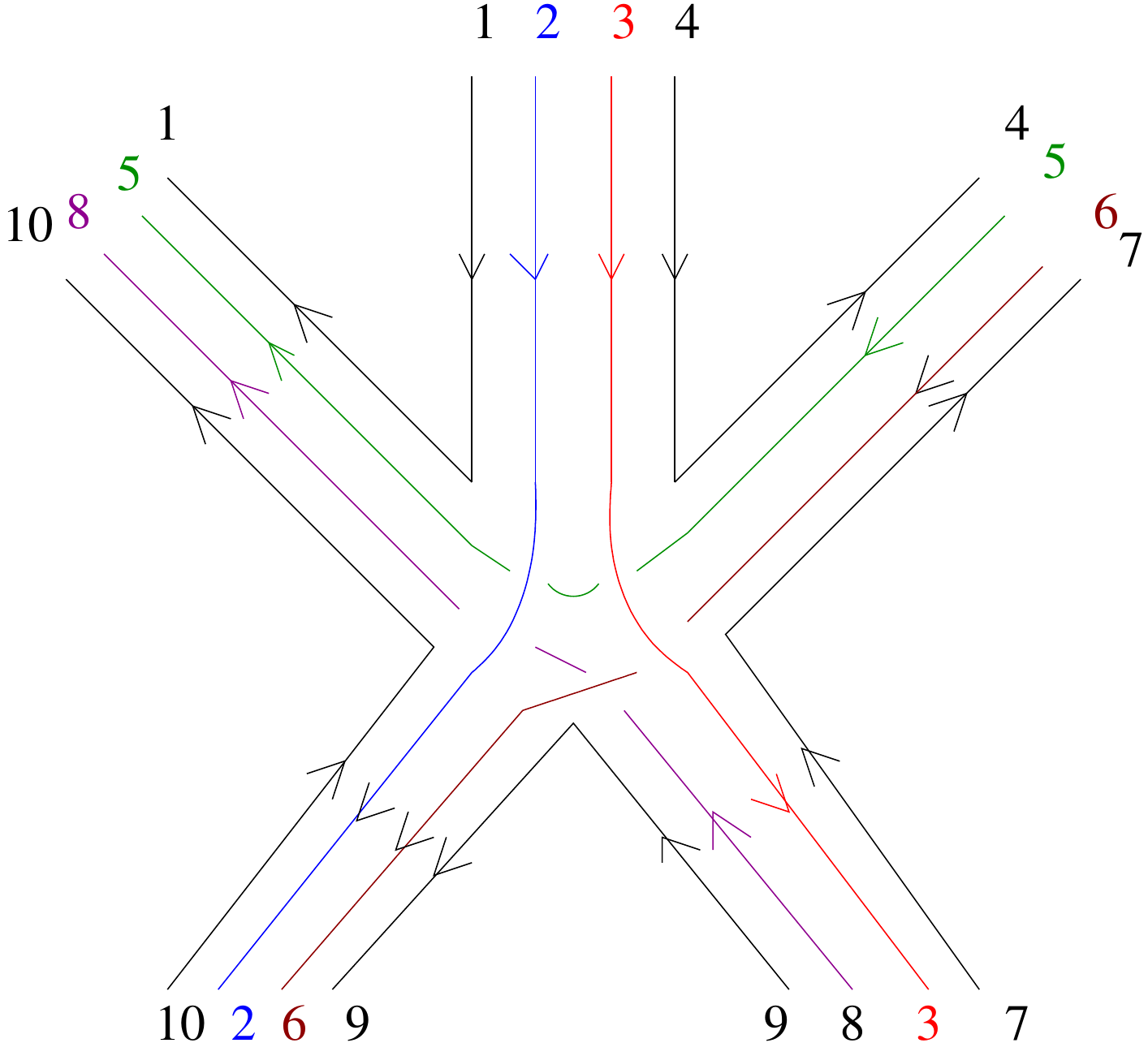,width=5cm} }
\caption{A simplicial vertex of a $4-$dimensional GFT. We have chosen here a particular matching and
orientation for each of the strands.}
\label{vertex}
\end{figure}

\subsection{Propagators}

We consider only field theories in which the propagator $C$ is Hermitian.
It can be considered either as an Hermitian operator $\phi \to C \phi $ acting on fields
or as its Hermitian kernel $C(g_1, \ldots, g_D; g'_1, \ldots, g'_D)$:
\be [ C\phi ] (g_1, \ldots , g_D) = \int  \dd g'_1  \ldots \dd g'_D C(g_1, \ldots, g_D; g'_1, \ldots, g'_D) \phi (g'_1, \ldots , g'_D) .
\ee
The corresponding normalized Gaussian measure of covariance $C$ is noted $d \mu_C$.
Hence
\be C(g_1, \ldots, g_D; g'_1, \ldots, g'_D)  = \int  \phi (g_1, \ldots , g_D) \phi (g'_1, \ldots, g'_D) d\mu_C .
\ee

\subsection{Graphs}

Graphs are generated by gluing together propagators and vertices, according to Wick contractions,
hence to Feynman rules.

\begin{definition}
A stranded graph is called {\it regular} if it has no {\it tadpoles} (hence any line $\ell$ joins two distinct vertices)
and no {\it tadfaces} (hence each face $f$ goes at most once through any line of the graph).
\end{definition}

It is convenient to introduce orientations on both lines and faces of stranded graphs.
Regular oriented graphs are natural since they are conveniently
described by two matrices

\begin{figure}
\centerline{\epsfig{figure=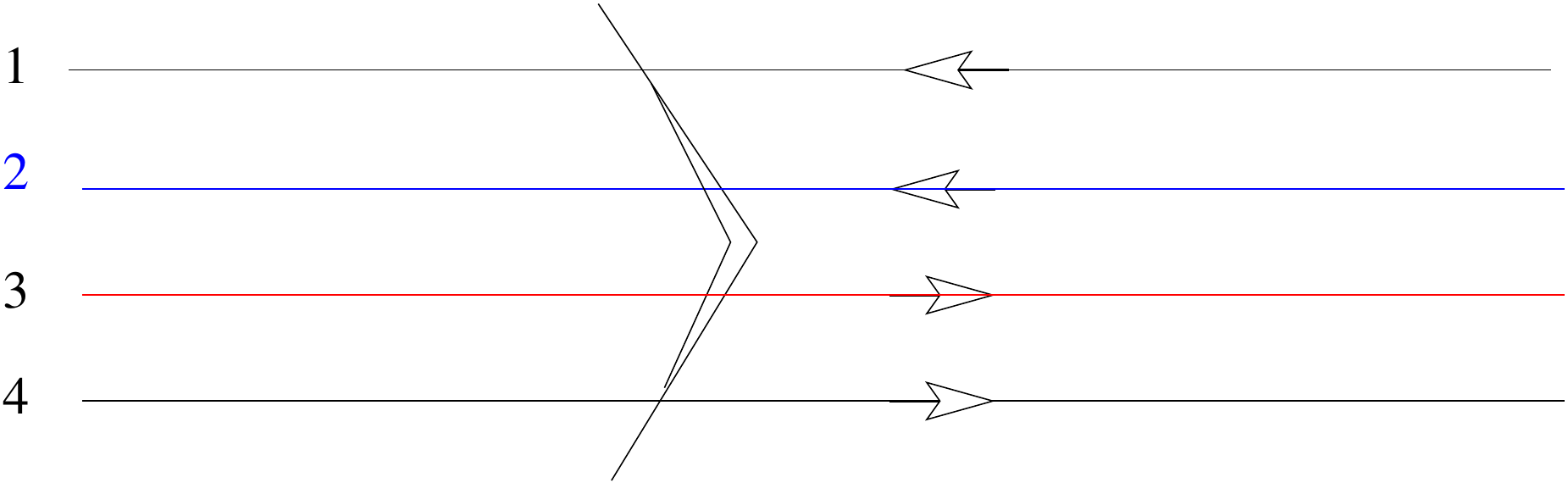,width=7cm} }
\caption{A stranded propagator with particular orientation; two strands have
$\eta_{\ell f}=1$ and the other two have $\eta_{\ell f}=-1$.}
\label{propa-fig}
\end{figure}

\begin{itemize}
\item the ordinary incidence matrix $\epsilon_{v, \ell}$ which has value $+1$ if the edge $\ell$ enters the vertex $v$,
$-1$ if the edge $\ell$ exits vertex $v$ and 0 otherwise. Hence $\sum_v \vert \epsilon_{v, \ell} \vert  = 2$ for each $\ell$.

\item the incidence matrix $\eta_{\ell, f}$ between faces and edges, which has value $+1$ if the face $f$
goes through edge $\ell$ in the same direction, $-1$ if the face $f$
goes through edge $\ell$ in the opposite direction
and 0 otherwise. Hence $\sum_f \vert \eta_{\ell,f} \vert  =D$ for each $\ell$ (see Fig. \ref{propa-fig}).
\end{itemize}

These orientations are useful to write down the {\it integrand} of the Feynman amplitudes.
However the {\it integrals}, that is the spin-foam amplitudes themselves,
do not depend on these orientations, at least
for the class of theories considered in this paper.

\begin{figure}
\centerline{\epsfig{figure=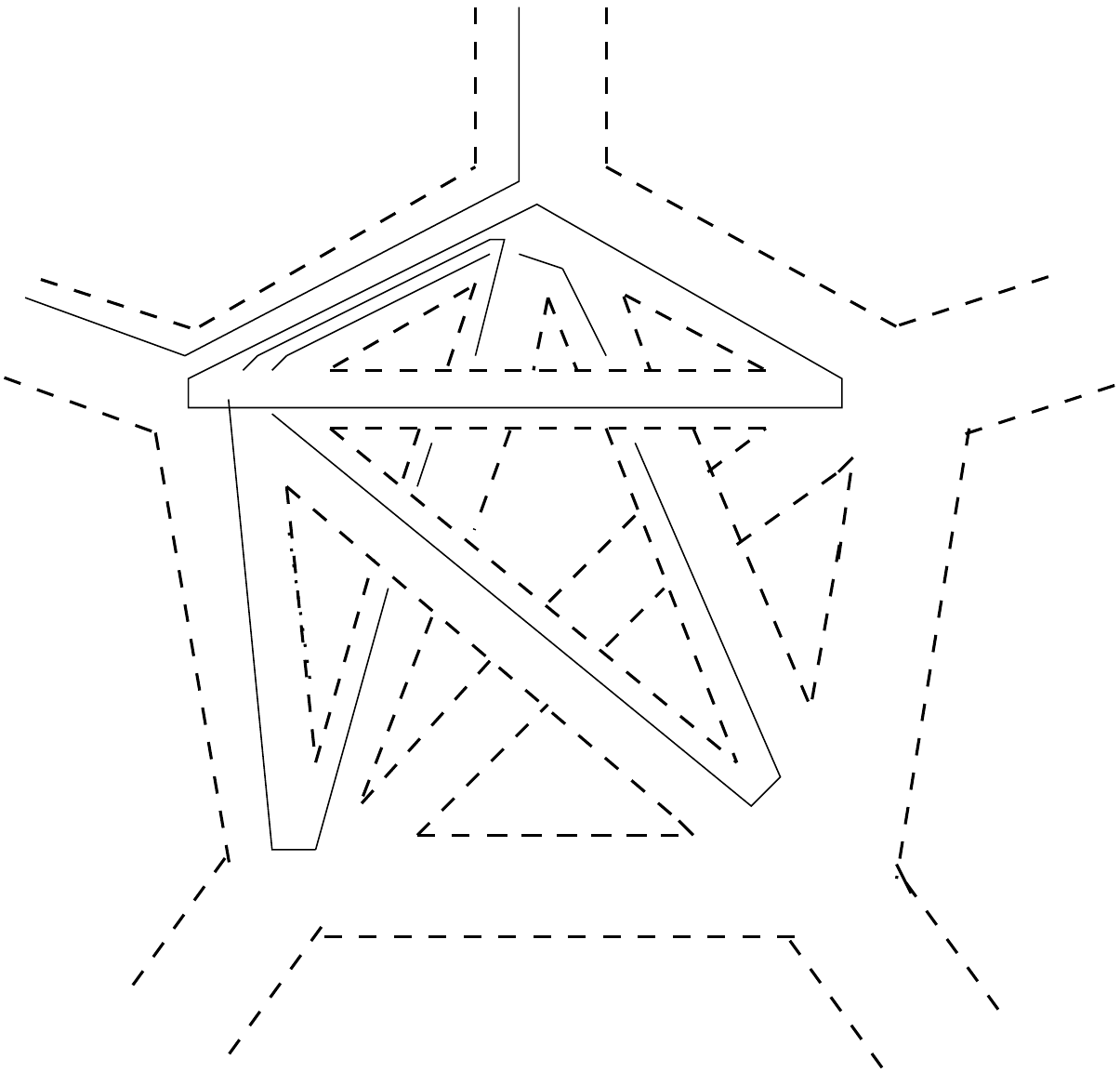,width=8cm} }
\caption{The ``starfish'' graph, quantum correction to the vertex. The dashed lines represent the edges (they do not correspond to strands). Each edge contains $4$ strands, there are $40$ such strands, forming $10$ closed faces and $10$ open faces. We have shown $4$ faces: $3$ closed and one open which take into account all $4$ strands of one particular edge (upper left).}
\label{5P-fig}
\end{figure}

From now on we consider only amplitudes for {\it regular} graphs. This is for convenience, as generalization to any graph of our formulas is possible. It has been argued that GFT should in fact be restricted to colored models, which generate only regular graphs \cite{gurau1,gurau2,gurau3}. Remark that every colorable stranded graph is regular, but the converse is not true; colorable graphs in particular have all their faces of even length, hence the ``starfish'' graph of Figure \ref{5P-fig}  with ten faces of length 3, although regular, is not colorable.

\subsection{The BF theory}

\subsubsection{Propagator in direct space}

By direct space we mean the representation which uses the group elements.

In the case of the BF theory the propagator\footnote{Beware that this propagator
is called $C$ in \cite{MNRS,Ben1}.}, here noted $\jj$, is just the projection on gauge invariant fields:
\be
\jj (\phi)(g_1, \ldots, d_D)=\int \dd h \; \phi(g_1 h,\ldots, g_D
h),
\ee
where the integral is performed over the group $SO(D)$ with respect to its Haar measure.
Let us remark that $\jj ^2=\jj$ so that the only eigenvalues are $0$ and $1$
(which means that the BF theory has no dynamics).
The operator $\jj$ is Hermitian with kernel
\be \label{kerneldirect1}
\jj (g_1, ..., g_D ; g'_1, ... g'_D)=\int \dd h \prod_{i=1}^D \delta (g_i h (g'_i)^{-1}).
\ee

\subsubsection{Amplitudes in direct space}

Suppose that we choose an arbitrary orientation of the lines
and faces of a graph $\mathcal G$ (which for simplicity has no external legs).

Combining together the vertices (\ref{vertexdirect}) (or the
generalization to dimension $D$) and the propagators
(\ref{kerneldirect1}) of the graph, the integration over all $g$
variables can be explicitly performed, leading  to the direct space
representation of the BF Feynman amplitude as an integral over line
variables $h$:

\begin{equation} A_{\mathcal G} =
\int \prod_{\ell \in {L}_{{\mathcal G}} } \dd h_\ell \prod_{  f  \in
{F}_{{\mathcal G}}} \delta \left(  {\vec {\prod}_{\ell \in f}
h_\ell^{\eta_{\ell f} }}   \right),
 \label{amp2}
\end{equation}
where ${L}_{{\mathcal G}}$, ${F}_{{\mathcal G}}$ are the set of
lines and faces of ${\mathcal G}$, respectively. The oriented
product $\vec{\prod}_{l \in  f} h^{\eta_{\ell f}}$ means that the
product of the variables $h_\ell$ has to be taken in the cyclic
ordering corresponding to the face orientation (starting anywhere on
the cycle).

As announced, the amplitude (\ref{amp2}) neither depends on the
arbitrary orientation of the lines, nor on those of the faces. A
pedestrian way to see this is to exploit carefully the parity of the
$\delta$ function and of the Haar measure under $h \to h^{-1}$.
Beware that  formula (\ref {amp2}) may be formal as this amplitude
can be infinite for many graphs.

\subsubsection{Amplitudes in the angular momentum basis}

The angular momentum basis uses the irreducible unitary representation
spaces $V^j$ of dimension $\rd_j\equiv 2j+1$. In this space there is a standard decomposition of unity
\be\label{ident-m}
{\mathbf 1}_j = \sum_m |j,m \ket \bra j,m|,
\ee
where $|j,m\ket, m\in[-j,j]$ is the usual orthonormal basis in $V^j$.

In dimension 3, using the Peter-Weyl theorem we can transform  {\eqref{amp2}}
to get the representation:

\begin{equation}
 A_{\mathcal G} =  \prod_f  \sum_{j_f}  d_{j_f}
 \prod_v  \{ 6 j \},  \label{amp3}
\end{equation}
where the Ponzano-Regge vertex is the $6j$ symbol (cfr appendix A)
corresponding to the 6 face indices meeting at the vertex.

In dimension 4 for simplicity we work with the covering group $SU(2)\times SU(2)$ of  $SO(4)$
 and we decompose the group elements as $g= (g_+ , g_-)$, with $
g_{\pm}\in SU(2)$. Moreover we write ${j}\equiv(j_+,j_-)$ for the
eigenvalues of the angular momentum $J$ in each $SU(2)$ component.
The Peter-Weyl decomposition leads to the similar angular momentum
representation:

\begin{equation}
 A_{\mathcal G} =  \prod_f\sum_{j_{f+} , j_{f-}}
d_{j_{f+}} d_{j_{f-}}  \prod_v  \{15 j_+ \} \{15 j_-  \}. \label{amp4}
\end{equation}
where $\{15 j \}$ are the 15j symbols (see for example \cite{PRS}
 for the definition and normalization conventions).
\subsubsection{Coherent states}

Consider $R^{(j)}{}^{m}_{\;\; k}(g)$, the matrix element of the
group element $g$ in the representation $j$, computed between the
states $\bra j,m|$ and  $|j, k \ket$. We have
\be\label{ident-coh}
{\mathbf 1}_j = \rd_j \sum_{mm'} |j,m \ket \bra j,m'|
\int_{{\rm SU}(2)} dg \, R^{(j)}{}^{m}_{\; \; j}(g) \overline{R}^{(j)}{}_{j}^{\; m'}(g) =
\rd_j \int_{{\rm SU}(2)} dg \, |j,g \ket \bra j,g|,
\ee
where we have introduced the notation:
\be\label{def}
|j,g \ket \equiv g |j,j\ket = \sum_m  |j,m \ket R^{(j)}{}^m_{\;\; j}(g) .
\ee
The states $|j,g \ket$ are the coherent states  \cite{perelomov},
and the last expression in (\ref{ident-coh}) is a decomposition of
the identity in terms of these coherent states.

Let us recall that the decomposition of the identity (\ref{ident-coh}) can be further simplified and
taken over the coset $G/H, G={\rm SU}(2), H={\rm U}(1)$:
\be
{\mathbf 1}_j = \rd_j \int_{G/H = S_2} dn \, |j,n \ket \bra j,n| \label{idn}
\ee
with $|j,n \ket=g_n|j,j\ket $ and $g_n$ defined in \eqref{gnrot}. We
suppress the domain of integration $G/H$ in what follows.

The states $|j,n\ket$ form a generating set in $V^j$ sometimes called ``overcomplete basis''.

Let us now turn to the  coherent states of the group $SU(2)\times
SU(2)$.
In fact
one has four possible
such coherent states which are given by acting with the same group
element $(g_+,g_-)\in SU(2)\times SU(2)$  on either of the following
states:
\be
|j,j\ket\otimes |j,j\ket, \quad|j,j\ket \otimes |j,-j\ket,
\quad|j,-j\ket \otimes |j,j\ket, \quad|j,-j\ket \otimes |j,-j\ket.
\ee
Note that these four states can be obtained from one another by the action of
an $SO(4)$ group element. However, if one considers only the action of the diagonal $\SU(2)$
subgroup of elements of the form $(g,g)$ then there are two inequivalent
states that cannot be related by such a transformation.

\subsubsection{The BF propagator and amplitudes using coherent states}

To prepare for the EPRL/FK propagator we  rewrite the BF propagator
inserting the coherent state decomposition of identity on each
strand.  Let us consider $SU(2)$ BF first. Since $\jj^2 = \jj$ we
can introduce two distinct $SU(2) $ gauge-averaging  variables, $u$
and $v$ at both ends of the propagator, instead of the single
variable $h$ (e. g.  $u$ on the side where $\epsilon_{v, \ell}=-1$ and
$v$ on the side where $\epsilon_{v, \ell}=+1$). Between these two
variables we insert the partition of unity (\ref{idn}). This does
not modify the propagator. Working out the algebra, we find
\bea
\jj (g ; g')
=\int \dd u \dd v
\prod_{f=1}^4 \sum_{j_{f}}  \rd_{j_{f}}
 \tr_{V_{j_{f}}} \left( u g_f (g'_f)^{-1} v^{-1}
{\mathbf 1}_{ j_f}\right), \label{genprop1}
\eea
with $g_f , g'_f, u$ and $v$ elements of $SU(2)$.
The index $f$ labels the four strands of the propagator, which belong to four different faces (since we consider only regular graphs).

To write down the amplitudes we need to introduce some notations.
There are now group variables $2 \vert { L_{\mathcal G}}\vert $,
$u_{\ell}$ and $v_\ell$, and $D\vert { L_{\mathcal G}}\vert $
variables $n$. The amplitude is again factorized over faces:
\begin{equation} A_{\mathcal G} =
\int \prod_{\ell \in {L}_{{\mathcal G}} } \dd u_\ell \dd v_{\ell}
\prod_{  f  \in {F}_{{\mathcal G}}}  {\mathcal A}_f .
 \label{amp8}
\end{equation}
To write down ${\mathcal A}_f$, let us number the vertices and lines
in the (anti)-cyclic order along a face $f$ of length $p$ as $
\ell_1, v_1 \cdots \ell_p, v_p$, with by definition $\ell_{p+1} =
\ell_1$. We have then
\bea \label{amp9}
\mathcal{A}_{f} =\sum_j d_j^{p+1} \int\prod_{a=1}^p d n_{\ell_a f} <j,
n_{\ell_a f} |h_{\ell_a,  v_a  }^{\eta_{\ell_a f}} h_{\ell_{a +1, v_a}}^{\eta_{\ell_{a+1}  f}}|j,
n_{\ell{a +1} , f}>,
\eea
where
\bea \label{ha}
h_{\ell_a , v_a}&=&v_{\ell_a}\;\; \textrm{if}\;\; \epsilon_{v_a,
\ell_a} =+ 1 \nn\\ h_{\ell_a , v_a}&=&u_{\ell_a} \;\;
\textrm{if}\;\; \epsilon_{v_a , \ell_a } = - 1.
\eea

\subsubsection{The $D=4$ BF case}

In $D=4$ we work with $SU(2)\times SU(2)$, the covering group of $SO(4)$;
we have the similar system $|j_+,n_+  \ket \otimes |j_-,n_-  \ket$ of coherent states
and the partition of unity on the space $V_j$, with $j= (j_+, j_-)$
\be
{\mathbf 1}_{j_+}\otimes {\mathbf 1}_{j_-}  =  {\mathbf 1}_j = \rd_{j_+} \rd_{j_-}\int \dd n_+\dd n_- |j_+,n_+\ket\otimes
|j_-,n_-\ket \bra j_+,n_+| \bra j_-,n_-|  . \label{TBF}
\ee
The gauge-averaging  variables, $u=(u_+,u_-)$
and $v=(v_+,v_-)$ at both ends of the propagator
are now elements of  $SU(2) \times SU(2)$.
Between these two variables we insert the partition of unity (\ref{TBF}) and we find
\bea
\jj (g ; g')
=\int \dd u \dd v
\prod_{f=1}^4 \sum_{j_{f+},j_{f-}}  \rd_{j_{f+}}
\rd_{j_{f-}} \tr_{V_{j_{f+}}\otimes
V_{j_{f-}}} \left( u g_f (g'_f)^{-1} v^{-1}
{\mathbf 1}_{ j_f}\right) \label{genprop}
\eea
with $g_f , g'_f, u$ and $v$ elements of $SU(2) \times SU(2)$, and we have
formulas similar to \eqref{amp8}-\eqref{amp9} for the amplitudes.

\subsection{The EPRL/FK GFT}

The EPRL/FK model introduces a modification of the propagator of the BF model,
while the vertex remains the same.
The EPFL/FK propagator has a structure similar to \eqref{genprop} but with
replacement of ${\mathbf 1}_j$ by a non-trivial projector. We notice at this point
that since this projector does not commute with $\jj$, it is not possible
to recombine $u$ and $v$ in a single gauge averaging variable $h$.

It implements in two steps the Plebanski constraints
with a non-trivial value of the Immirzi parameter $\gamma$. Starting
from the (\ref{genprop}) expression of the $BF$ propagator in the coherent states representation,
the first step adds the constraint $j_+/j_- = (1+\gamma) /(1- \gamma) $
on the representations summed.
Remark however that this equation may have no solution
(e. g. if $\gamma$ is irrational) and should be true only in an asymptotic sense in the ultraspin
limit where $j_+$ and $j_-$ are both very large.

More precisely this constraint reads
\bea
\gamma>1 &&\qquad \qquad j_{\pm}=  \frac{\gamma \pm 1}{2} j,
\;\;\; \; \; \qquad  n_+=n_- \label{imm-1}\\
\gamma<1 &&\qquad \qquad j_{\pm}=  \gamma_{\pm} j = \frac{1\pm
\gamma}{2} j, \;\;\;n_+=n_-,  \label{imm-2}.
\eea
where $j_{\pm}, j$ are half  integers.
\footnote{Moreover, in the case $\gamma>1$ the coherent states to be used
below are the ones in their ``anti-parallel''
version $|j,n\ket\otimes \overline{|j,{n}\ket}$ \cite{FreKra}.}

From now on we consider only the case $0 < \gamma \le 1$
where the EPRL and  FK models coincide. At $\gamma=1$, the EPRL/FK model reduces to a single $SU(2)$ BF theory (see below).

The second step replaces in each strand of (\ref{genprop}) the identity ${\mathbf 1}_j$ by
a projector $T_j^{\gamma}$ whose definition is
\be
T^{\gamma}_{ j}=   d_{j_+ + j_-} \bigl[  \delta_{j_{f-}/j_{f+} = (1-\gamma) /(1+ \gamma) }  \bigr] \int \dd n  |j_+,n \ket\otimes
|j_-,n \ket \bra j_+,n | \otimes \bra j_-, n | \label{TBFEPRL} .
\ee
Let us notice here that, in the angular momentum basis, the operator
$T^\gamma_j$ takes the form
\bea
T^\gamma_j&=&\sum_{k,\tilde k, m,\tilde m} \bigl(j_+,k;j_-,\tilde
k|j_++j_-,k+\tilde k\bigr) \bigl(j_++j_-,m+\tilde m|j_+,m;j_-,\tilde
m\bigr)\nn\\
&& |j_+k>\otimes|j_-\tilde k\rangle\langle  j_+m|\otimes<j_-\tilde
m|\; \delta_{m+\tilde m, k+\tilde k}, \label{Tangmom}
\eea
where $(.|.)$ denotes the Clebsh-Gordan coefficients.

Grouping the four strands of a line defines a $\TT$ operator that
acts separately and independently on each strand of the propagator:
\be  \TT = \oplus_{j_{f}}    \otimes_{f=1}^4
 T_{j_f}^{\gamma}
\ee
so that the EPRL/FK propagator is
\bea
\label{defC}
&&C= \jj \TT \jj ;\ \ \
C(g,g')=  \int \dd u \dd v
\prod_{f=1}^4\sum_{j_f} \bigl[  \delta_{j_{f-}/j_{f+} = (1-\gamma) /(1+ \gamma) }  \bigr]
\alpha_{j_f} \beta_{j_f}\int dn_f \\
&&{\mathrm{Tr}}_{j_{f+}\otimes j_{f-}} \left( u g_f \;
(g^{'}_f)^{-1}v^{-1} |j_{f+},n_f>\otimes |j_{f-},n_f\rangle \langle
j_{f+},n_f|\otimes<j_{f-},n_f| \right),\nonumber
\eea
where
\bea
\alpha_{j}=d_{j_+}d_{j_-},\ \ \
\beta_j = d_{j_{+} + j_{-}}
\eea

\begin{lemma}
The operator $C$  is Hermitian.
\end{lemma}
\noindent{\bf Proof}
We have\bea
\tr_{j_{f+}\otimes j_{f-}} \left(u g_f (g'_f)^{-1 } v^{-1}
T^\gamma_{j_f}\right)&=
& \beta_{j_f} \int\dd n \tr_{j_{f+}\otimes j_{f-}} \\
&&\left(u \epsilon^T\epsilon g_f \epsilon^T\epsilon (g'_f)^{-1
}\epsilon^T\epsilon v^{-1} n_f|j_{f+}j_{f-}\rangle\langle
j_{f+}j_{f-}|n_f^\dag\right), \nn
\eea
where we have inserted the product $\epsilon^T\epsilon = 1$
with $\epsilon\in SU(2)$ defined by (\ref{epsilonop}).
We arrive at
\bea
\tr_{j_{f+}\otimes j_{f-}} (u g_f (g'_f)^{-1 } v^{-1}
T^\gamma_{j_f})&=&\beta_{j_f}\int\dd n \tr_{j_{f+}\otimes j_{f-}} (u
\epsilon^T \bar g_f  (g'_f)^{T }\epsilon v^{-1}
n_f|j_{f+}j_{f-}\rangle \langle j_{f+}j_{f-}|n_f^\dag\nn)\\
&=&
\beta_{j_f}\int\dd n \tr_{j_{f+}\otimes j_{f-}} (
\epsilon^T \bar u \bar g_f  (g'_f)^{T } v^{T}\epsilon
n_f|j_{f+}j_{f-}\rangle \langle j_{f+}j_{f-}|n_f^\dag\nn )\\
&=&\beta_{j_f}\int\dd n \tr_{j_{f+}\otimes j_{f-}} (\epsilon \bar
n_f |j_{f+}j_{f-}\rangle \langle j_{f+}j_{f-}| n_f^T\epsilon^T v
g'_f g_f^{-1} u^{-1} ) \nn \\&=&\tr_{j_{f+}\otimes j_{f-}} (v g'_f
g_f^{-1 } u^{-1} T^\gamma_{j_f}),
\eea
which implies the Lemma. \hfill QED

\medskip

Since the propagator is hermitian, Feynman amplitudes are again independent
of the orientations of faces and propagators.

\begin{lemma}
$\TT$ is a projector, namely $(\TT)^2= \TT$.
\end{lemma}
\noindent{\bf Proof} In the coherent states basis it is easier to
check that $(\TT)^3= (\TT)^2$, which adding Hermiticity of $\TT$
implies the Lemma. The equation $(\TT)^3= (\TT)^2$  follows from the
same equation on each strand, since
\bea
&& \beta^3_j \int dn dn' dn''  |j_{+},n> \otimes\vert  j_{-},n
\rangle\langle j_+ + j_- , n \vert j_+ + j_- ,n'>
 \nn\\
&&<j_+ +j_- , n' \vert j_+ +j_- ,n''  \rangle \langle j_{+}, n'' \vert  \otimes\langle  j_{-}, n''|  \nn \\
&=& \beta^2_j \int dn dn''  |j_{+}, n \rangle  \otimes \vert j_{-} ,
n \rangle \langle j_+ + j_- , n \vert
 j_+ +j_- ,  n''  \rangle \langle j_{+} ,n'' \vert \otimes j_{-} , n''|,
\eea
where we have used that ${\mathbf 1}_{j_+ + j_-} = \beta_j \int dn'
\vert j_+ + j_- , n'\rangle \langle  j_+ + j_- ,   n' \vert  $.
\hfill QED

Let us also notice that the lemma is easily proven in the angular
momentum basis, where, from \eqn{Tangmom} it easily follows that
$(T^\gamma_j)^2=T^\gamma_j$.

 Since $\TT$ and $\jj$ do not commute,
the propagator $C$ can have non-trivial {\it spectrum} (with
eigenvalues between 0 and 1). Slicing the eigenvalues should allow a
renormalization group analysis. This is why we would like to call
this kind of theories {\it dynamic} GFT's.

Remark that since $\TT$ is a projector, the propagator $C$ of the EPRL/FK theory
is bounded in norm by the propagator of the $BF$ theory, and that Feynman amplitudes
for the EPRL/FK theory are therefore {\it bounded} by those of the $BF$ theory; in particular
we expect milder ultraspin (large $j$) divergences in EPRL/FK.

\subsubsection{Amplitudes}
Combining the propagator and the vertex expressions, the
integrations over all $g,g'$ group variables can be performed
explicitly, leading to the amplitude of any  graph ${\mathcal G}$.
This amplitude is  given by an integral of a product over all faces
of the graph as in \eqref{amp8}, but the amplitudes for faces are
different.

To compute these  face amplitudes we distinguish between
closed faces (no external edges) and open faces (which end on external edges).

Using the same numbering of the  $p$ edges and vertices along a closed face,
its amplitude is given by
\bea
\mathcal{A}_f=\int \prod_{a=1}^p \bigl(\dd g_{\ell_a} \dd g'_{\ell_a}
\bigr) \sum_{j_{\ell_a}} \alpha_{j_{\ell_a}}
 \tr_{j_{\ell_a} +\otimes j_{\ell_a} -} \bigl( (u_{\ell_a}  g_{\ell_a}
(g'_{\ell_a})^{-1}v_{\ell_a}^{-1})^{\eta_{\ell_a f}}
T_{j_{\ell_a}}^\gamma \bigr) \prod_v V_v,
\label{faceampl0}
\eea
where the constraint on $j_+, j_-$ is implicitly understood from now
on.
 We can
perform the $g$ integrals using \eqref{gg-1} or \eqref{gg} and we
arrive at
\be
\mathcal{A}_f =\sum_{j_f} \alpha_{j_f} \tr_{j_{f+}\otimes j_{f-}}
\overrightarrow{\prod}_{\stackrel{a=1}{}}^p \left(h_{\ell_a,v_a}
^{\eta_{\ell_a f}}
 h_{\ell_{a+1}, v_{a}}^{\eta_{\ell_{a+1}f}}
T_{j_f}^\gamma\right),
 \label{faceampl}
\ee
with $h_{\ell_a,v_a}$ defined in \eqn{ha} and $\ell_{p+1}=\ell_1$.
Note that we use \eqref{gg-1} or \eqref{gg} to take into account the
fact that $\eta_{\ell f}$ can change when we follow a face $f$. We
 find
\bea
\mathcal{A}_f &=&\sum_{j_f} \alpha_{j_f}\int\prod_{a=1}^p \beta_{j_f} d n_{\ell_a, f}
 \langle j_{f+} n_{\ell_{a},f}  |  h_{\ell_a,v_a, +}^{\eta_{\ell_a f}}
 h_{\ell_{a+1},v_a, +}^{\eta_{\ell{a+1} f}}|j_{f+} n_{\ell_{a+1},f}\rangle \nn \\
&&\times\langle j_{f-} n_{\ell_{a},f}  |h_{\ell_a,v_a,
-}^{\eta_{\ell_a f}}
 h_{\ell_{a+1},v_a, -}^{\eta_{\ell{a+1} f}}
|j_{f-}  n_{\ell_{a+1},f}\rangle. \label{facecoher}
\eea

\subsubsection{BF limit}

Let us see how we recover the $SU(2)$ BF model  in the limit $\gamma=1$. In this limit
$j_-=0$, hence $j_+=j_++j_-$.
Thus we are
left with
\bea
\mathcal{A}_{f} |_{\gamma=1}=\sum_{j_f} d_{j_f}^{p+1}
\int\prod_{a=1}^p d n_{\ell_a, f}  \langle j_f,  n_{\ell_a, f}
|h_{\ell_a, v_a , +}^{\eta_{\ell_a f}}  h_{\ell_{a+ 1},v_a,
+}^{\eta_{\ell_{a+1} f}}|j_f, n_{\ell_{a+1}, f}\rangle,
\eea
where only one $SU(2)$ copy appears. We can  use
the completeness relation for coherent states \eqn{idn}, and
cyclicity of the trace to reorder the product according to lines instead of vertices and we
obtain
\bea
\mathcal{A}_{f} |_{\gamma=1} =\sum_j d^{2}_j\int d n \langle j
n|\prod_{\stackrel{a=1}{}}^p h_{\ell_a ,v_a ,  +}^{\eta_{\ell_a f}}
 h_{\ell_{a}, v_{a+1} , +}^{\eta_{\ell_{a} f}}   |j n\rangle.
\eea
Redefining
$t_{\ell_a}=u_{\ell_a , +}^{-1} v_{\ell_a , +}$ we finally obtain
\be
\mathcal{A}_{f}|_{\gamma =1} =\sum_j d_j^2\int d n \langle j n|\vec
\prod_{ a \in f} t_{\ell_a} ^{\eta_{\ell_a f}}|j n\rangle=
\delta(\vec\prod_{\ell\in f} t_\ell^{\eta_{\ell f}}),
\label{faceoog}
\ee
consistently with \eqn{amp2}.

\subsubsection{Amplitudes with  external edges}

For a face with external edges the expression is slightly modified,
as there is no integration on the external data.

Let us call $G, \tilde G$ the group labels of the incoming and
outgoing  external strands respectively.  We omit the edge index
$\ell$ in the following.  Moreover we indicate with
$u_{\textrm{in}}, v_{\textrm{in}}, u_{\textrm{out}},
v_{\textrm{out}}$ the gauge transformations on the incoming,
outgoing edges respectively.  Let $q$ be the number of internal
strands. The expression of the resulting face amplitude is similar
to \eqn{faceampl0} except for the fact that we don't integrate on
the external labels. On chosing
\be
\eta_{{\textrm{in}}\,f}=\eta_{{\textrm{out}}\,f}=1
\ee
that is,  the incoming and outgoing strand oriented according to the
face,  we find using \eqref{gg-1} and  \eqref{gg}
\bea
\mathcal{A}_{ext}= \sum_{j}\alpha_j\times \tr\Bigl[ u_{\textrm{in}}
G\tilde G^{-1} v_{\textrm{out}}^{-1}T^\gamma_j
 u_{\textrm{out}}u_{\ell_1} T^\gamma_j
\left(\prod_{a=2}^{q}u_{\ell_a}
 v_{\ell_{a+1}}T^\gamma_j\right) v_{\ell_{q}}^{-1} v_{\textrm{in}}^{-1}T^\gamma_j\Bigr] \label{faceampl3}
\eea
where, to simplify the notation, we have chosen all propagators
oriented according to the face. It is immediately verified that it
 reduces to \eqn{faceampl} with
$p=q+2$ if we glue together the external edges with the insertion of
a delta function  $\delta(G \tilde G^{-1}) {\delta}_{\sigma
\sigma'}$.

\section{Stationary phase for BF and EPRL/FK models}

Let $\cG$ be a graph in a GFT corresponding to the BF or EPRL/FK models, made of  $V$ vertices,
$L$ edges and $F$ faces, usually labelled by letters $v$, $\ell$ and $f$.  In the coherent state basis,
its amplitude can in general be written as
\begin{equation}
{\cal A}_{\cG}=\sum_{j_{f} \le \Lambda}{\cal N}\int \prod dh \prod dn
\exp\big\{\sum_{f}j_{f}S_{f}[h,n]\big\},\label{amplisaddle}
\end{equation}
where ${\cal N}$ is a normalization factor which is a rational function of the spins. As explained in
the previous sections, the precise form of the face action and of the number of group variables $h\in\mbox{SU(2)}$ and unit vectors $n\in S^{2}$ depends on the choice of the model. Note that the sums over the spins $j_f$ may lead to divergences, so that we introduce an ultraspin cut-off $\Lambda$ that restricts the summation to spins below $\Lambda$. To derive the superficial power counting, we set $j_{f}=jk_{f}$ with $k_{f}\in[0,1]$ and use the stationary phase method to derive the large $j$ behavior of
\begin{equation}\label{kffactor}
\int \prod dh \prod dn\exp\big\{j\sum_{f}k_{f}S_{f}[h,n]\big\} .
\end{equation}
If the action is complex but has a negative real part, the contribution to this integral are quadratic fluctuations around zeroes of the real part of $S$ which are stationary points of its imaginary part,
otherwise the integral is exponentially suppressed as $j\rightarrow\infty$.

\subsection{BF models in the coherent state representation}

\label{BF}

For BF models we have one group element $h_{l}\in \mbox{SU(2)}$. In dimension 4 one should work with $\mbox{SU(2)}\times\mbox{SU(2)}$ instead, which lead to two independent copies of the previous amplitude, so that we restrict ourselves to $\mbox{SU(2)}$ for simplification. The amplitude is given
by (\ref{amp9}). Including the $k_f$ factor of \eqref{kffactor} in the action and using
\begin{equation}
\langle n,j|g|n',j\rangle=\langle n|g|n'\rangle^{2j},
\end{equation}
with $|n\rangle$ a shorthand for $|\frac{1}{2} n\rangle$,
 it can be written in the form \eqref{amplisaddle} with
\begin{equation}
S_{f}[h,n]=2k_{f}\log\langle n|\mathop{\prod}\limits_{\ell\in\partial f}^{\longrightarrow}
h_{\ell}^{\eta_{\ell,f}}|n\rangle
\end{equation}
Note that $|n\rangle\langle n|$ is a projector
\begin{equation}
|n\rangle\langle n|={\textstyle \frac{1}{2}}\big(1+\sigma\cdot n\big),
\end{equation}
so that the action reads
\begin{equation}
S_{f}[h,n]=k_{f}\log\mbox{Tr}\Big[\big(\mathop{\prod}\limits_{\ell\in\partial f}^{\longrightarrow}h_{\ell}^{\eta_{\ell,f}}\big)\big(1+\sigma\cdot n\big) \Big].\label{actionBF}
\end{equation}
Since the action is the logarithm of the trace of the product of an unitary element and a projector, it is clear that its real part is negative (it is the logarithm of the modulus of the trace, obviously bounded by 1) and maximal when the unitary element is one. This is attained at $h_{\ell}=1$, but other solutions may be possible. In particular, the BF amplitude is invariant under the gauge transformations $g_{v}$ at any vertex
\begin{equation}
h_{\ell}\rightarrow g_{v}h_{\ell}g_{v'}^{-1}
\end{equation}
for any edge from the vertex $v$ to the vertex $v'$. Therefore gauge transformations of the trivial solution $h_{\ell}=1$ yield other equivalent solutions. More generally, there is a continuum of solutions connected to the trivial one which will translate into flat directions in the saddle point approximation.

To perform the saddle point expansion, we expand the group element to second order as
\begin{equation}
h_{\ell}=1-\frac{A_{\ell}^{2}}{2}+\mathrm{i}\,A_{\ell}\cdot\sigma+O(A_{\ell}^{3}),
\end{equation}
with $A\in su(2)\times su(2)$ a Lie algebra element. By the same
token, we expand the unit vectors as
\begin{equation}
n_{f}=n_{f}^{(0)}+\xi_{f}-\frac{\xi_{f}^{2}}{2}n^{(0)}_{f}+O(\xi_{f}^{3}),\qquad\mbox{with}\quad n_{f}^{(0)}\cdot\xi_{f}=0.
\end{equation}
This expansion is determined by the requirement that $n_{f}^{2}=1$ up to third order terms. To alleviate the notation, we drop the superscript $(0)$ in the sequel. Let us consider a face with edges $\ell_{1},\dots,\ell_{p}$, then to second order
\begin{equation}
\mathop{\prod}\limits_{\ell\in\partial f}^{\longrightarrow}h_{\ell}^{\eta_{\ell,f}}
=1-\frac{A_{f}^{2}}{2}+\mathrm{i}\,\sigma\cdot A_{f}-\mathrm{i}\,\sigma\cdot\Phi_{f}
\end{equation}
with
\begin{equation}
A_{f}=\sum_{1\leq a\leq p}\eta_{\ell_{a},f}A_{\ell_{a}}\qquad\mbox{and}\qquad
\Phi_{f}=\sum_{1\leq a<b\leq p}\eta_{\ell_{a},f}\eta_{\ell_{b},f} \, A_{\ell_{a}}\wedge A_{\ell_{b}}.
\end{equation}
Expanding then to second order \eqref{actionBF}, we get
\begin{equation}
S_{f}[A_{\ell},\xi_{f}]=2k_{f}\Big\{\mathrm{i}\,n_{f}\cdot A_{f}-\frac{A_{f}^{2}}{2}+\frac{(n_{f}\cdot A_{f})^{2}}{2}+\mathrm{i}\,\xi_{f}\cdot A_{f}+\mathrm{i}\,n_{f}\cdot\Phi_{f}\Big\}
\end{equation}
and we have to estimate
\begin{equation}
\int\prod_{\ell}dA_{\ell}\prod_{f}d\xi_{f}\,\exp 2j
\sum_{f} k_{f}\Big\{\mathrm{i}\,n_{f}\cdot A_{f}-\frac{A_{f}^{2}}{2}+\frac{(n_{f}\cdot A_{f})^{2}}{2}+\mathrm{i}\,\xi_{f}\cdot A_{f}+\mathrm{i}\,n_{f}\cdot\Phi_{f}\Big\}
\end{equation}
as $j\rightarrow\infty$. Note that we do not integrate over the vectors $n_{f}$, the latter have to be chosen so that they are extrema of the imaginary part of $S$. Because all terms except the first one $\sum_{f}k_{f}n_{f}\cdot A_{f}$ are of second order, the imaginary part is stationary if and only if
\begin{equation}
\sum_{f} \mathrm{i}\,k_{f}n_{f}\cdot A_{f}=\sum_{\ell,f} \mathrm{i}\,\eta_{\ell,f}k_{f}n_{f}\cdot A_{\ell}=0
\qquad\forall A_{\ell}\in{\Bbb R}^{3},
\end{equation}
which amounts to the closure condition
\begin{equation}
\sum_{f}\eta_{\ell,f}k_{f}n_{f}=0,\label{closure}
\end{equation}
to be fulfilled for any edge $\ell$.  This is the well-known requirement that, in the semi-classical limit,  the vectors $j_{f}n_{f}$ are the sides of a triangle (resp. the area bivectors of a tetrahedron) that propagates along $\ell$ in dimension 3 (resp. dimension 4).
The solutions of the closure conditions range from non-degenerate to maximally degenerate. In three dimensional (resp. four dimensional ) BF theory,  a solution is said to be non-degenerate if all the tetrahedra (resp. 4-simplices) corresponding to the vertices of the graph have
maximal dimension. At the opposite end, we say that a solution is maximally degenerate if all the vectors $n_{f}$ are proportional to a single one $n_{0}$,
\begin{equation}
n_{f}=\sigma_{f}n_{0}\qquad\mbox{with}\qquad \sigma_{f}\in\left\{-1,+1\right\}.
\end{equation}

\subsubsection{Maximally degenerate case}
Let us first concentrate on the maximally degenerate solutions and show that for simply connected graphs (i.e. every closed loop can be shrunk to a point by deforming it through the faces), the quadratic saddle point approximation yields
an upper bound estimate
\begin{equation}
{\cal A}_{\cG}\lesssim \Lambda^{3F-3r}
\end{equation}
with $r$ the rank of the $L\times F$ incidence matrix $\eta_{\ell,f}$. This is in accordance with the general results for BF theory presented , see also \cite{BonSme}.

To derive this result, we proceed with the following five steps.

\begin{enumerate}
\item
For a maximally degenerate solution, the closure constraints amount to
\begin{equation}
\sum_{f}\eta_{\ell,f} x_{f}=0
\end{equation}
with $x_{f}=k_{f}\sigma_{f}$. Since the rank of the matrix $\eta_{\ell,f}$ is $r$, the $x_{f}$ live in a $F-r$ dimensional subspace. The signs $\sigma_{f}$ have to be adjusted so that $k_{f}>0$. We end up with a summation over $F-r$ independent spins in \eqref{amplisaddle}. Let us note that since the incidence matrix has integer coefficients, all the spins may always be chosen to be half-integers, after multiplication by a suitable integer.

\item
For BF theory in the coherent state representation, we have a factor of $d_{j}^{2}$ per face, so that the normalization behaves as
\begin{equation}
{\cal N}=(d_{j})^{2F}\sim j^{2F},
\end{equation}
where we have discarded an inessential multiplicative constant as $j\rightarrow\infty$.

\item
The integration over $\xi_{f}$ can be performed using the Fourier representation of the $\delta$ function
\begin{equation}
\int d\xi_{f}\,\exp\left\{\mathrm{i}\,j\xi_{f}\cdot A\right\}=\frac{1}{j^{2}}\delta_{{n_{0}}^{\perp}}(A_{f})
\end{equation}
with an inessential factor of $(2\pi)^{2}$ absorbed in the integration measure. Note that the vector $\xi_{f}$ is constrained to lie in the plane orthogonal to $n_{f}$, so that it enforces the constraint $A_{f}=0$ only in that plane. Since $A_{f}=\sum_{\ell}\eta_{\ell,f}A_{\ell}$, these constraints are not independent.
The number of independent constraints  is  $2r$, since everything takes place in the plane orthogonal to a vector $n_{f}=\sigma_f n_{0}$ which does not depend on the face. Altogether, the integration over the $\xi_{f}$ yield a factor of $j^{-2r}$ and implement the constraints
\begin{equation}
\sum_{\ell}\eta_{\ell,f}A_{\ell}=0\qquad\mbox{in the directions orthogonal to }n_{0}.\label{constraintdegenerate}
\end{equation}
\item
Using the previous constraints, the real part of the action involving $A$ only vanishes, $(A_{f})^{2}-(n_{f}\cdot A_{f})^{2}=0$.

\item
Because the graph is simply connected, the constraints \eqref{constraintdegenerate} imply the existence of vectors $C_{v}\in{\Bbb R}^{3}$ attached to the vertices and orthogonal to $n_{0}$ such that
\begin{equation}
A_{\ell}-(n_{0}\cdot A_{\ell})n_0=C_{s(\ell)}-C_{t(\ell)},\label{solutionconstraint}
\end{equation}
with $s(\ell)$ (resp. $t(\ell)$) the source (resp. the target) of the edge $\ell$. Then, the phase associated to a face $f$ reads
\begin{equation}
n_{0} \cdot \Phi_{f}=\sum_{\ell\in\partial f}\eta_{\ell,f}\,n_{0}\cdot(C_{s(\ell)}\wedge C_{t(\ell)}).
\end{equation}
The total contribution of all the faces to the action vanishes since
\begin{equation}
\sum_{f}k_{f}n_{f} \cdot \Phi_{f}=\sum_{f,\ell}\eta_{\ell,f}k_{f}\,n_{f}\cdot(C_{s(\ell)}\wedge C_{t(\ell)})=\sum_{\ell}\Big(\sum_{f}\eta_{\ell,f}k_{f}n_{f}\Big)\cdot\Big(C_{s(\ell)}\wedge C_{t_{\ell}}\Big)=0,
\end{equation}
using the closure condition \eqref{closure}. 
\item
However, its is important to note that the components of $A$ parallel to $n_{0}$ are not constraint by \eqref{solutionconstraint} and their contribution to the action vanishes identically in the quadratic approximation.  This is the reason why we only get an upper bound in the maximally degenerate case.

\end{enumerate}
Accordingly, the bound for the amplitude can be estimated as
\begin{equation}
\sum_{F-r\,\mbox{\tiny independent spins }\atop
\mbox{\tiny of order }j\leq \Lambda}j^{2F}\times j^{-2r}\sim\Lambda^{3F-3r}\label{degree}
\end{equation}
which is the result obtained in \cite{linhom}.

It is interesting to note that for a simply connected graph, the rank $r$ of the incidence matrix can be written as $r=F-(V-1)$. Indeed, the system of equations \eqref{constraintdegenerate}, whose rank is $2r$ allows to write the $2L$ variables $A_{l}$ in terms of $2(V-1)$ differences $C_{v}-C_{v'}$, all in the direction orthogonal to $n_{0}$. Therefore, one has $2L-2r=2V-2$, so that $r=L-V+1$ and the amplitude of a simply connected graph scales as
\begin{equation}
{\cal A}_{\cG}\quad\lesssim\quad \Lambda^{3(\chi_{\cG}-1)},
\end{equation}
with $\chi_{\cG}=F-L+V$ the Euler characteristics of the graph. This also reproduces the result of \cite{BonSme}, since $\chi_{\cG}=\mbox{dim}H^{2}_{\cG}-\mbox{dim}H^{1}_{\cG}+\mbox{dim}H^{0}_{\cG}
=\mbox{dim}H^{2}_{\cG}+1$ for a simply connected graph. This is also in accordance with the results of \cite{linhom} for graphs with planar jacket: The faces $F_{\mbox{\tiny jacket}}$ of the planar jacket obey $F_{\mbox{\tiny jacket}}-L+V=2$, since the associated surface has the topology of a sphere, and the remaining faces are in bijections with the cycles followed by the  $N=F-F_{\mbox{\tiny jacket}}$ strands in the middle, so that the degree of divergence reads
$\omega_{\cG}=3(F_{\mbox{\tiny jacket}}-L+V-1)+3N=3(N+1)$.

\subsubsection{Non-degenerate case}

In the non-degenerate case, the situation is slightly more complicated. The integration over the variables $\xi_{f}$ yields a system of equations analogous to \eqref{constraintdegenerate}, but now with a vector $n_{f}$ that varies from face to face,
\begin{equation}
\sum_{\ell}\eta_{\ell,f}A_{\ell}=0\qquad\mbox{in the direction orthogonal to }n_{f}. \label{constraintnondegenerate}
\end{equation}
Because for fixed $\ell$ the three (resp. four) vectors $\eta_{\ell,f}n_{f}$ in dimension 3 (dimension 4) span a space of dimension 2 (resp. 3), all the components of $A_{\ell}$ appear in the system \eqref{constraintnondegenerate}, contrary to the maximally degenerate case \eqref{constraintdegenerate}, which only involves the components of $A_{\ell}$ orthogonal to $n_{0}$.  In the example treated
in detail below (see Sect. \ref{self}), \eqref{constraintnondegenerate} turns out to be equivalent to $\sum_{\ell}\eta_{\ell,f}A_{\ell}=0$, which has rank $3r$ and yields the same degree of divergence. In the general case, we expect non-degenerate configurations to have a less divergent behavior, since the degree of divergence $3F-3r$ obtained in \cite{linhom} in the Abelian case is expected to be an overestimate in the general case and is the correct asymptotic behavior at least for many graphs.

\subsubsection{Two dimensional case}

To close this section, let us see how the saddle point method allows us to recover the results presented in \cite{Baez} in the simplest case of  BF theory in dimension 2. In this case GFT graphs are ordinary ribbon graphs with trivalent vertices representing triangles. The closure condition reads $k_{f_{1}}n_{f_{1}}=k_{f_{2}}n_{f_{2}}$ for every edge that separates two different faces and is vacuous for edges that appear twice as we go along a face (if we restrict ourselves to triangulations of orientable surfaces).
Thus, for a genus 0 graph there is a single spin $j$ and a single unit vector $n$. Moreover, the graph is simply connected so the phase disappears. Choosing an arbitrary face, all other $F-1$ constraints are independent since the corresponding faces span a sphere with one hole that can be filled with the remaining face. Thus, $r=F-1$ and $\omega=3F-3(F-1)=3$, as expected from the relation (see \cite{Baez})
\begin{equation}
{\cal A}_{\cG}=\sum_{j\leq \Lambda}\,j^{\chi_{\cG}}\sim\Lambda^{\chi_{\cG}+1},\label{Euler}
\end{equation}
with $\chi_{\cG}$ the Euler characteristic.

At higher genera the graph is no longer simply connected and  the contribution of the antisymmetric part may be decisive. For instance, for the non-planar double tadpole $\cG_1$ (torus topology)
\begin{equation}
A_{\cG_1}=\int dh_{1}dh_{2}\,\delta(h_{1}h_{2}(h_{1})^{-1}(h_{2})^{-1})
\end{equation}
an expansion to second order yields the action
\begin{equation}
S[n,A_{1},A_{2}]=\mathrm{i}\,\,jn\cdot A_{1}\wedge A_{2}
\end{equation}
The rank of this quadratic form is $4$ and the Gau\ss ian integration over $A_{1}$ and $A_{2}$ yields
\begin{equation}
{\cal A}_{\cG_1}=\sum_{j\leq \Lambda}\,j^{2}\times j^{-4/2}\sim\Lambda,
\end{equation}
in accordance with $\eqref{Euler}$.

\subsection{Divergence of the self-energy in the EPRL/FK model}
\label{self}

\begin{figure}
\centerline{\epsfig{figure=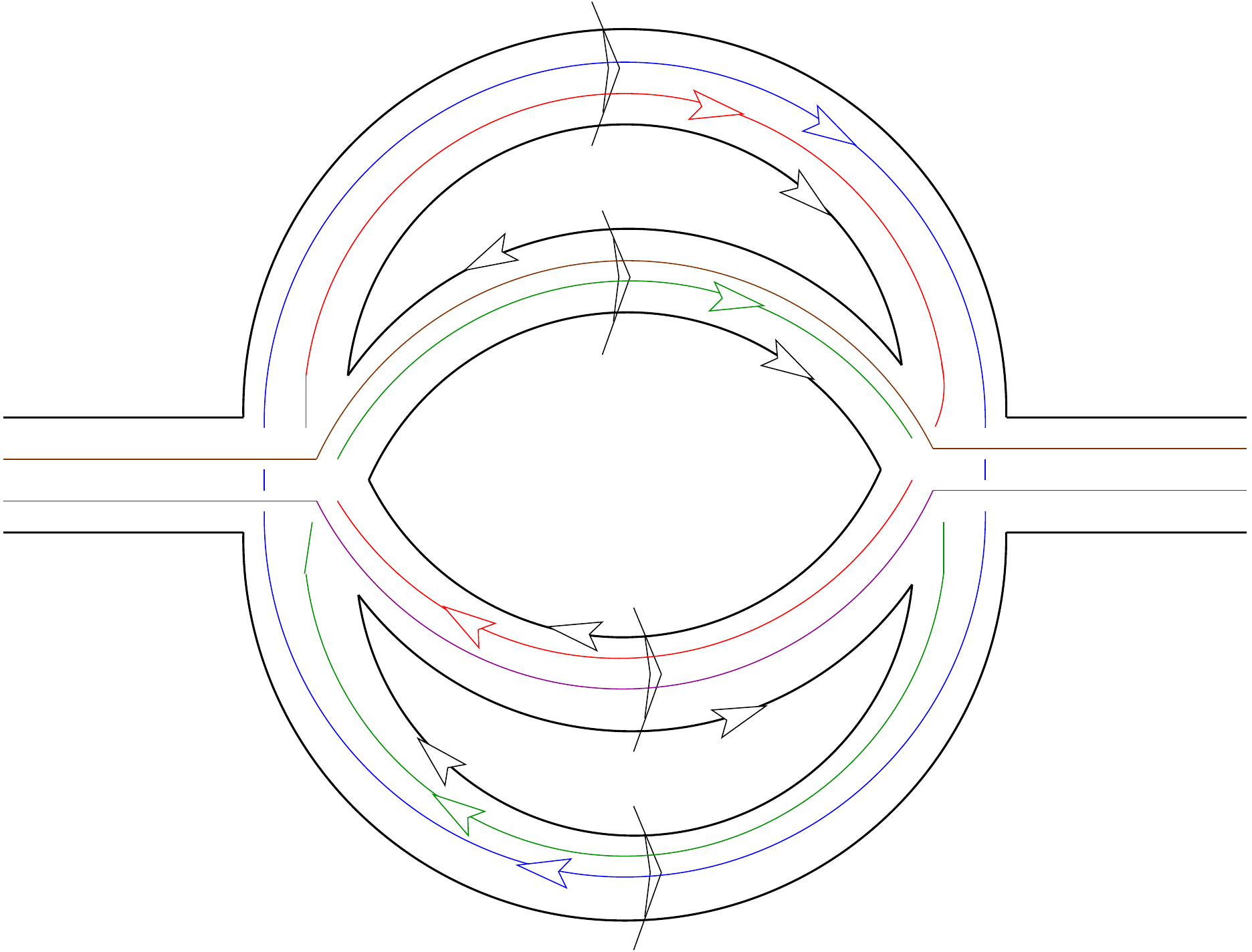,width=9cm} }
\caption{The ``self-energy'' graph $\cG_2$, quantum correction to the propagator.}
\label{2P-fig}
\end{figure}

The  self-energy graph $\cG_2$ of Fig. \ref{2P-fig} has $4$ open
faces. It has $6$  closed faces with two edges each. We label the
internal propagators with an index $a$ ranging from  $1$ to  $4$ and
orient them in the same direction. We label the $6$ closed faces
with pairs of indices $(a,b)$, $a<b$. Its amplitude reads
\bea
{\cal A}_{\cG_2}&=&\prod_a \dd u_a^{\pm} {\dd v^a}^{\pm}\prod_{a<b}
\mathcal{A}_{ab}
\eea
where, from \eqn{facecoher} the face amplitude reads
\bea
\mathcal{A}_{ab}&=&\sum_jd_{j_+} d_{j_-} \beta_j^2 \int d n_{ab} d
n_{ab}'  \langle j_+ n_{ab}|u_{a +} u_{b +}^{-1}|j_+ n_{ab}'\rangle
\langle j_+ n_{ab}'|v_{b +} v_{a +}^{-1}|j_+ n_{ab}\rangle
\nn\\
&\times&\;\;\;\;\langle j_- n_{ab}|u_{a -} u_{b -}^{-1}|j_-
n_{ab}'\rangle\langle j_- n_{ab}'|v_{b -} v_{a -}^{-1}|j_-
n_{ab}\rangle.
\eea
Using \eqref{aj2}, we rewrite the amplitude above as
\bea
\mathcal{A}_{ab}&=&\sum_j d_{j_+} d_{j_-}\beta^2_j\int d n d n'
(\langle n|u_{a +} u_{b +}^{-1}|n'\rangle\langle n'|v_{b +} v_{a
+}^{-1}| n\rangle)^{2j_+}
\nn\\
&\times&\;\;\;\;(\langle  n|u_{a -} u_{b -}^{-1}| n'\rangle\langle
n'|v_{b -} v_{a -}^{-1}| n\rangle)^{2j_-}.\label{a2cycle}
\eea
In order to perform a stationary phase analysis we rewrite the graph
amplitude as
\bea
{\cal A}_{\cG_2} &=&\sum_{j_{f}}\int
\prod_{a}du_{a}^{\pm}\,\prod_{a}dv_{a}^{\pm}\,
\prod_{i}dn_{i}\,\prod_{f}\Big\{(d_{j_{f}})^{2}d_{j^{+}_{f}}d_{j^{-}_{f}}
\exp\big\{jS^{+}_{f}+jS^{-}_{f}\big\}\Big\} \label{self-energy}
\eea
with $j_{f}^{\pm}=j\gamma^{\pm}k_{f}$, $k_{f}\in[0,1]$ and $j$
large. There is one coherent state per strand, which amounts here to
label the coherent states by a couple of a face and an edge
$i=(f,l)$ such that $\eta_{l,f}\neq 0$. The face action for $f=ab$
can be written as
\begin{align}
S^{\pm}_{f}=
2\gamma^{\pm}k_{f}\log\big\{\langle n_{f,a}|u_{a}^{\pm}(u_{b}^{\pm})^{-1}|n_{f,b}\rangle\langle n_{f,b}|v_{b}^{\pm}(v_{a}^{\pm})^{-1}|n_{f,a}\rangle\big\}.
\end{align}

\begin{figure}
\centerline{\epsfig{figure=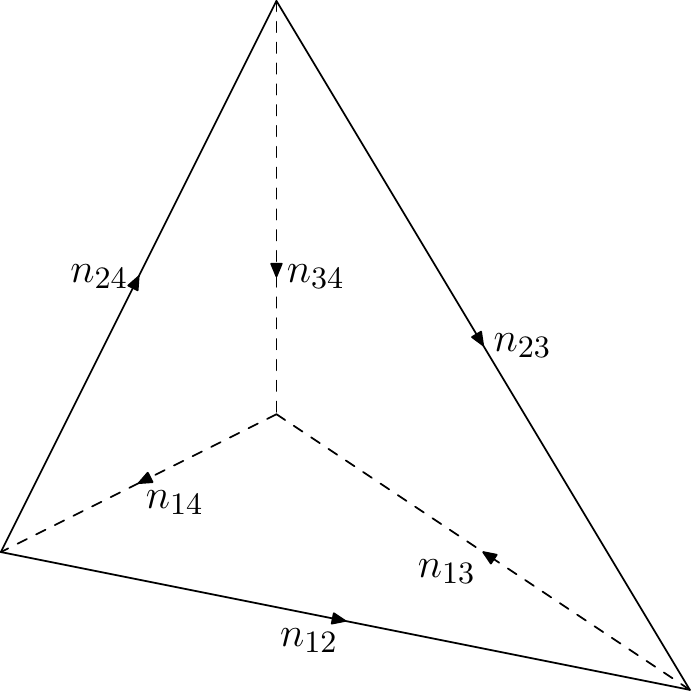,width=7cm} }
\caption{The tetrahedron illustrating the closure condition.}
\label{tetra}
\end{figure}

We employ the saddle point technique around the identity and develop the group elements as follows
\begin{equation}
u_{a}^{\pm}=1-\frac{(A_{a}^{\pm})^{2}}{2}+\mathrm{i}\,\sigma\cdot A_{a}^{\pm}+O(A_{a}^{\pm})^{3},\qquad
v_{a}^{\pm}=1-\frac{(B_{a}^{\pm})^{2}}{2}+\mathrm{i}\,\sigma\cdot B_{a}^{\pm}+O(B_{a}^{\pm})^{3} .
\end{equation}
Morover, introducing the projector
\begin{equation}
|n_{i}\rangle\langle n_{i}|=\frac{1+\mathrm{i}\,\sigma\cdot n_{i}}{2},\label{projector}
\end{equation}
the action at the identity for the face $f=ab$ reads
\begin{align}
S^{\pm}_{f}[1,1,n_{i}]=
\gamma^{\pm}k_{ab}\log\mbox{Tr}\Big\{\frac{1+\mathrm{i}\,\sigma\cdot n_{f,a}}{2}\frac{1+\mathrm{i}\,\sigma\cdot n_{f,b}}{2}\Big\}=
\gamma^{\pm}k_{ab}\log\Big\{\frac{1+n_{f,a}\cdot n_{f,b}}{2}\Big\},
\end{align}
which is negative except for $n_{f,a}=n_{f,b}=n_{f}$. Therefore, we perform the expansion of the coherent state around an unit vector common to all the strands of the face
\begin{equation}
n_{i}=n_{f}+\xi_{i}-\frac{(\xi_{i})^{2}}{2}n_{f}+O(\xi_{i})^{3},\qquad \mbox{with}\qquad n_{f}\cdot\xi_{i}=0,
\end{equation}
otherwise the integral is exponentially damped. To perform the expansion at second order of the action, it is convenient to rewrite this action as
\begin{align}
S^{\pm}_{f}=&
\gamma^{\pm}k_{f}\log\mbox{Tr}\big\{|n_{f,a}\rangle\langle n_{f,a}|u_{a}^{\pm}(u_{b}^{\pm})^{-1}|n_{f,b}\rangle\langle n_{f,b}|\big\}\cr
&+\gamma^{\pm}k_{f}
\log\mbox{Tr}\big\{|n_{f,b}\rangle\langle n_{f,b}|v_{b}^{\pm}(v_{a}^{\pm})^{-1}|n_{f,a}\rangle\langle n_{f,a}|\big\}\cr
&-\gamma^{\pm}k_{f}\log\mbox{Tr}\big\{|n_{f,a}\rangle\langle n_{f,a}||n_{f,b}\rangle\langle n_{f,b}|\big\} .
\end{align}
Using the projector \eqref{projector}, the expansion to second order only involves traces of products of Pauli matrices and  is straightforward but rather lengthy. A crucial intermediate result is the expansion to second order in $A_{1},A_{2},\xi_{1},\xi_{2}$,
\begin{eqnarray}
&&\frac{1}{4}
\mbox{Tr}\Big\{\big[1+\sigma\big(n+\xi_{1}-\frac{(\xi_{1})^{2}}{2}n\big)\big]
\big[1-\frac{(A_{1}-A_{2})^{2}}{2}+\mathrm{i}\,\sigma\cdot\big(A_{1}-A_{2}+A_{1}\wedge A_{2}\big)\big] \nonumber\\
&&
\big[1+\sigma\big(n+\xi_{2}-\frac{(\xi_{2})^{2}}{2}n\big)\big]\Big\}
\ =\ 1-\frac{(A_{1}-A_{2})^{2}}{2}+\mathrm{i}\,n\cdot\big(A_{1}-A_{2}+A_{1}\wedge A_{2}\big)\nonumber\\
&&-\frac{\big(\xi_{1}-\xi_{2}\big)^{2}}{4}
+\mathrm{i}\,\big(\xi_{1}+\xi_{2}\big)\frac{A_{1}-A_{2}}{2}+\big(\xi_{1}-\xi_{2}\big)\frac{n\wedge\big(A_{1}-A_{2}\big)}{2} .
\end{eqnarray}
Gathering all terms together and taking the logarithm, we get
\begin{eqnarray}
S^{\pm}_{f}[A^{\pm},B^{\pm},\xi_{i}]&=&k_{f}\gamma^{\pm}\Big\{-(A^{\pm}_{a}-A^{\pm}_{b})^{2}+\big[n_{f}\cdot(A^{\pm}_{a}-A^{\pm}_{b})\big]^{2}
\\
&&-(B^{\pm}_{b}-B^{\pm}_{a})^{2}+\big[n_{f}\cdot(B^{\pm}_{b}-B^{\pm}_{a})\big]^{2}\nonumber\\
&&+\mathrm{i}\,n_{f}\cdot\big(A^{\pm}_{a}-A^{\pm}_{b}+B^{\pm}_{b}-B^{\pm}_{a}\big)+\mathrm{i}\,n_{f}\cdot\big(A^{\pm}_{a}\wedge A^{\pm}_{b}+B^{\pm}_{b}\wedge B^{\pm}_{a}\big)\nonumber\\
&& +\mathrm{i}\,\big(\xi_{f,a}+\xi_{f,b}\big)\cdot\big(A^{\pm}_{a}-A^{\pm}_{b}+B^{\pm}_{b}-B^{\pm}_{a}\big)\nonumber\\
&&-\frac{\big(\xi_{f,a}-\xi_{f,b}\big)^{2}}{2}+\big(\xi_{f,a}-\xi_{f,b}\big)\cdot\big[n_{f}\wedge\big(A^{\pm}_{a}-A^{\pm}_{b}-(B^{\pm}_{b}-B^{\pm}_{a})\big)\big]\Big\} . \nonumber
\end{eqnarray}
To complete the computation, one has to perform a Gau\ss ian integration with an action $S=\sum_{f}(S_{f}^{+}+S_{f}^{-})$. In order to disentangle this computation, it is convenient to perform the following change of variables
\begin{equation}
A^{\pm}_{a}=A_{a}\pm \gamma^{\mp}X_{a}\qquad\mbox{and}\qquad
B^{\pm}_{a}=B_{a}\pm \gamma^{\mp}Y_{a}.\label{changeself}
\end{equation}
The interest of this change of variables is that the terms linear in $A^{\pm}$ and $B^{\pm}$ now only involve $A$ and $B$, while in the quadratic terms, the pair of variables $A$ and $B$ on one side and the pair $X$ and $Y$ on the other side decouple. We shall return in greater detail to this change of variable in section \ref{change} in the case of a arbitrary graph, since it allows to separate the action, at the level of the quadratic approximation, into a SU(2) BF action (variables $A$ and $B$) and an ultralocal potential that only involves uncoupled variables attached to the vertices (variables $X$ and $Y$). Turning back to the self-energy, we get
\begin{align}
S_{f}[A,B,X,Y,\xi]&=S^{+}_{f}[A^{+},B^{+},\xi]+S^{-}_{f}[A^{-},B^{-},\xi]\cr
&=k_{f}\Big\{-(A_{a}-A_{b})^{2}+\big[n_{f}\cdot(A_{a}-A_{b})\big]^{2}
-(B_{b}-B_{a})^{2}+\big[n_{f}\cdot(B_{b}-B_{a})\big]^{2}\cr
&\qquad+\mathrm{i}\,n_{f}\cdot\big(A_{a}-A_{b}+B_{b}-B_{a}\big)+\mathrm{i}\,n_{f}\cdot\big(A_{a}\wedge A_{b}+B_{b}\wedge B_{a}\big)\cr
& \qquad+\mathrm{i}\,\big(\xi_{f,a}+\xi_{f,b}\big)\cdot\big(A_{a}-A_{b}+B_{b}-B_{a}\big)\cr
&\qquad-\frac{\big(\xi_{f,a}-\xi_{f,b}\big)^{2}}{2}+\big(\xi_{f,a}-\xi_{f,b}\big)\cdot\big[n_{f}\wedge\big(A_{a}-A_{b}-(B_{b}-B_{a})\big)\big]\Big\}\cr
&+k_{f}\gamma^{+}\gamma^{-}\Big\{-(X_{a}-X_{b})^{2}+\big[n_{f}\cdot(X_{a}-X_{b})\big]^{2}+\mathrm{i}\,n_{f}\cdot\big(X_{a}\wedge X_{b}\big)\cr
&\qquad\qquad\qquad-(Y_{b}-Y_{a})^{2}+\big[n_{f}\cdot(Y_{b}-Y_{a})\big]^{2}+
\mathrm{i}\,n_{f}\cdot\big(Y_{b}\wedge Y_{a}\big)\Big\}.
\end{align}
Performing the Gau\ss ian integration over the two dimensional vector  $\chi_{f}=\xi_{f,a}-\xi_{f,b}$, one has
\begin{align}
&\int d\chi_{f}\, \exp jk_{f}\Big\{
- \frac{\chi_{f}^{2}}{2}+\chi_{f}\cdot\big[n_{f}\wedge\big(A_{a}-A_{b}-(B_{b}-B_{a})\big)\big]
\Big\}=\cr
&\qquad\frac{2\pi}{j_{f}}\,\exp\frac{jk_{f}}{2}\Big[n_{f}\wedge\big(A_{a}-A_{b}-(B_{b}-B_{a})\big)\Big]^{2} . \label{chiself}
\end{align}
Discarding an inessential constant in the limit $j\rightarrow\infty$ to alleviate the notations, the graph amplitude can therefore be written as
\begin{align}
{\cal A}_{\cG_2}=&\sum_{j_{f}}\,j^{18}\,\Big\{
\int \prod_{a}dA_{a}\,\prod_{a}dB_{a}\,\prod_{f}d\xi_{f}\exp jS_{BF}(A,B,\xi)\cr
&\qquad\times\int \prod_{a} dX_{a}\exp jQ(X)\times\int \prod_{a} dY_{a}\exp jQ(Y)\Big\},
\end{align}
with $\xi_{f}=\xi_{f,a}+\xi_{f,b}$. The  BF-like action is
\begin{align}
S_{BF}[A,B,\xi]&=\sum_{a<b}k_{ab}\Big\{-\frac{1}{2}\Big[n_{f}\wedge\big(A_{a}-A_{b}+B_{b}-B_{a}\big)\Big]^{2}\cr
&\qquad\qquad+\mathrm{i}\,n_{ab}\cdot\big(A_{a}-A_{b}+B_{b}-B_{a}\big)+\mathrm{i}\,n_{ab}\cdot\big(A_{a}\wedge A_{b}+B_{b}\wedge B_{a}\big)\cr
&\qquad\qquad+\mathrm{i}\,\xi_{ab}\cdot\big(A_{a}-A_{b}+B_{b}-B_{a}\big)
\Big\},
\end{align}
while the ultra local terms are
\begin{equation}
Q[X]=\gamma^{+}\gamma^{-}\sum_{a<b}k_{ab}\Big\{\big[n_{ab}\wedge(X_{a}-X_{b})\big]^{2}+\mathrm{i}\,n_{ab}\cdot\big(X_{a}\wedge X_{b}\big)
\Big\} .
\end{equation}
The Gau\ss ian integral over the variables $A$ and $B$ can be evaluated using the same techniques as in the section \ref{BF} devoted to BF theory. Firstly, there are four closure conditions, one for each edge $a$,
\begin{equation}
\sum_{b=a+1}^{4}k_{ab}n_{ab}-\sum_{b=1}^{a-1}k_{ba}n_{ba}=0\label{closureself}
\end{equation}
or explicitly,
\begin{equation}
\left\{
\begin{array}{rcl}
k_{12}n_{12}+k_{13}n_{13}+k_{14}n_{14}&=&0,\cr
-k_{12}n_{12}+k_{23}n_{23}+k_{24}n_{24}&=&0,\cr
-k_{13}n_{13}-k_{23}n_{23}+k_{34}n_{14}&=&0,\cr
k_{14}n_{14}+k_{24}n_{24}+k_{34}n_{34}&=&0 . \cr
\end{array}
\right.
\end{equation}
Note that the first three equations are independent while the last one is their sum, so that the rank of
the incidence matrix $\eta_{l, f}$ is 3. Let us investigate the case of non-degenerate configurations, which means
that the six vectors $k_{ab}n_{ab}$ span a three dimensional space. Geometrically, the solution of these closure
constraints can be realized by constructing a tetrahedron (see Fig.\ref{tetra}) with faces labelled 1,2,3,4 and
assigning the vector $k_{ab}n_{ab}$ to an edge between faces $a$ and $b$. Consequently, we sum over 6 independent
spins in  \eqref{self-energy}.

The Gau\ss ian integration over the variables $\xi_{ab}$ imposes the constraints
\begin{equation}
A_{a}-B_{a}=A_{b}-B_{b}\qquad\mbox{in the direction orthogonal to }n_{ab}.\label{constraintself}
\end{equation}
However, out of the 12 relations in \eqref{constraintself}, only 9 of them are independent and they are equivalent to
\begin{equation}
A_{a}-B_{a}=C\label{constraintselfequiv}
\end{equation}
with $C\in{\Bbb R}^{3}$. First of all, it is clear that any solution of \eqref{constraintselfequiv} is a solution of \eqref{constraintself}.  Let us show that the converse also holds. Let consider all the equations involving edge 1,
\begin{equation}
A_{1}-B_{1}=A_{a}-B_{a}\qquad\mbox{in the direction orthogonal to }n_{1a},\qquad a\in\left\{2,3,4\right\}
\end{equation}
Because of the closure constraint $\sum_{a>1}k_{1a}n_{1a}=0$, the relation  $A_{1}-B_{1}=A_{a}-B_{a}$ holds in the one dimensional space orthogonal to all  three vectors $k_{1a}n_{1a}$. Thus the vectors $A_{a}-B_{a}$ are all equal along this direction.  We then repeat the same reasoning for the other edges and conclude that the vectors $A_{a}-B_{a}$ are all equal along all directions using the non-degeneracy condition. As conclusion the rank of \eqref{constraintself} is 9 since it reduces the 12 degrees of freedom of the 4 vectors  $A_{a}-B_{a}$ to a single vector and the Gau\ss ian integration over $\xi_{ab}$ enforcing this constraint yields a factor of $j^{-9}$.

Using this constraint, the real part of the quadratic action obviously vanishes. The imaginary part can be dealt with using the techniques of section \ref{BF}. Using first the constraint, we write
\begin{equation}
0=(A_{a}+B_{b})\wedge (B_{a}+A_{b})=A_{a}\wedge B_{a}+B_{b}\wedge A_{b}+A_{a}\wedge A_{b}+B_{b}\wedge B_{a}.
\end{equation}
After summation over all faces, the net contribution of the phases to the amplitude vanishes
\begin{equation}
\sum_{a<b}\mathrm{i}\,k_{ab}n_{ab}\cdot\Big[A_{a}\wedge A_{b}+B_{b}\wedge B_{a}\big]=-
\sum_{a<b}\mathrm{i}\,k_{ab}n_{ab}\cdot\Big[A_{a}\wedge B_{a}+B_{b}\wedge A_{b}\big],
\end{equation}
where we have use the closure constraints. Altogether, the integration  over the variables $A$, $B$ and $\xi$ yields
\begin{equation}
\int \prod_{a}dA_{a}\,\prod_{a}dB_{a}\,\prod_{f}d\xi_{f}\exp jS_{BF}(A,B,\xi)\quad\sim\quad j^{-9}\label{ABself}
\end{equation}
as $j\rightarrow \infty$. Note that this, together with a $j^{12}$ arising form the coherent state representation of the $\delta$ function ($j^{2}$ per face) and a summation over 6 independent spins reproduces
\begin{equation}
\sum_{\mbox{\tiny 6 independent spins }\sim j\,<\Lambda}\, j^{12}\times j^{-9}\sim \Lambda^{9},
\end{equation}
which is the known result for SU(2) BF theory. Since the rank $r$ of the incidence matrix $\eta_{l, f }$ is 3, this reproduces with non-degenerate configurations the results of \cite{linhom}, with a degree of divergence $3F-3r$.

Let us now consider the Gau\ss ian integral over the independent variables $X_{a}$ and $Y_{a}$,
\begin{equation}
\int \prod_{a} dX_{a}\exp jQ(X)\quad\sim\quad j^{-\frac{\mbox{\tiny rank}(Q)}{2}}
\end{equation}
which simply amounts to compute the rank of the quadratic form
\begin{equation}
Q[X]=\gamma^{+}\gamma^{-}\sum_{a<b}k_{ab}\Big\{\big[n_{ab}\wedge(X_{a}-X_{b})\big]^{2}+\mathrm{i}\,n_{ab}\cdot\big(X_{a}\wedge X_{b}\big)
\Big\} .
\end{equation}
This quadratic form is associated with a symmetric bilinear form
\begin{equation}
B[X,Z]=\frac{1}{4}\Big(Q(X+Z,X+Z)-Q(X-Z,X-Z)\Big)
\end{equation}
and its kernel is defined as the subspace of the variables $X$ such that $B[X,Z]=0$ for all $Z$.
First, notice that $B[X,Z]$ is complex but the variables $X$ and $Z$ are real. Therefore, the orthogonality
condition  $B[X,Z]=0$ has to be fulfilled for the real and the imaginary part separately. Since the real part is
positive (but not definite positive), $X$ has to obey
\begin{equation}
\sum_{a<b}k_{ab}\big[n_{ab}\wedge(X_{a}-X_{b})\big]^{2}=0,
\end{equation}
or equivalently $X_{a}-X_{b}=0$ in the plane orthogonal $n_{ab}$.
Using the non-degeneracy of the configuration, an analysis identical
to that of the constraints \eqref{constraintself} leads to $X_{a}=C$
with $C\in{\Bbb R}^{3}$ that do not depend on the edge. Then, the
imaginary part of the relation $B(X,Z)=0$ reads
\begin{equation}
\sum_{a<b}\Big\{k_{ab}n_{ab}\cdot\big(C\wedge Z_{b}\big)+k_{ab}n_{ab}\cdot\big(Z_{a}\wedge C\big)\Big\}=0,
\end{equation}
which is identically fulfilled for any $Z_{b}$ because of the closure condition \eqref{closureself}.
Finally, the rank of $Q$ is the dimension of the orthogonal of its kernel, since the latter has dimension 3
and we have 4 vector variables $X_{a}$, we obtain $\mbox{rank}(Q)=12-3=9$, so that the Gau\ss ian integral
over $X$ yields an power of $j^{-9/2}$.  Obviously, the same holds for the integration over $Y$.

Therefore, we obtain the power counting for the self-energy with non-degenerate configurations as follows
\begin{equation}
\sum_{\mbox{\tiny 6 independent spins }\sim j\,<\Lambda}\, j^{24}\times j^{-6}\times j^{-9}\times \big(j^{-9/2}\big)^{2}
\sim \Lambda^{6},
\end{equation}
with the factor $j^{24}$ arising from a $d_{j^{+}}d_{j^{-}}\sim j^{2}$ for each of the 6 faces and a factor $d_{j}\sim j$
for each of the two strand in each face. The factor $j^{-6}$ results form the Gau\ss ian integration over the 6 variables
$\chi_{f}=(\xi_{f,a}-\xi_{f,b})$ in \eqref{chiself} and the $j^{-9}$ from the integration over $A$ and $B$ in
\eqref{ABself}. This reproduces the result of \cite{PRS}, with non-degenerate configurations. Note that this is an asymptotic behavior and not a mere bound as we had before, since all the zero modes of the quadratic approximation correspond to gauge degrees of freedom \eqref{constraintselfequiv}.

It is also of interest to notice that this result should also hold with finite non-zero spins on the external faces.
Indeed, since the latter remain finite, the contribution of the external faces to the action can be neglected
as $j\rightarrow\infty$.

Finally, let us mention that we have derived this power counting with non-degenerate configurations.
In the next section, we shall discuss maximally degenerate configurations.

\subsection{A bound for maximally degenerate configurations}

\label{change}

Consider a general graph $\cG$ in the EPRL/FK model with $F$ faces $f$. Since we are going to take the limit $j_{f}\rightarrow\infty$ for the internal spins, the contribution of the external
faces can be neglected, as long as their spins remain finite. Recall that the graph amplitude can be written as
\begin{equation}
{\cal A}_{\cG}=\int\,\prod dh\,\prod {\cal A}_{f},
\end{equation}
with the face amplitude given by \eqn{facecoher}.
The graph amplitude may be rewritten as in \eqn{amplisaddle}
\begin{equation}
{\cal A}_{f}=\sum_{j_{f}} \Big\{
d_{j^{+}_{f}}d_{j^{-}_{f}}\big(d_{j}\big)^{p}\int \prod dn\,\prod dh
\, \exp j\sum_{f}\big\{S_{f}^{+}+S_{f}^{-}\big\}\Big\}.
\end{equation}
 with
\begin{equation}
S_{f}^{\pm}[n,h]=2k_{f}\gamma^{\pm}\sum_{1\leq q\leq
p}\log\Big\{\langle n_{f,l_{q}}|
(h^{+}_{v_{q},l_{q}})^{{\displaystyle
\epsilon}_{v_{q},l_{q}}{\displaystyle\eta}_{l_{q},f}}(h^{+}_{v_{q},l_{q+1}})^{{\displaystyle\epsilon}_{v_{q},l_{q+1}}{\displaystyle\eta}_{l_{q+1},f}}|n_{f,l_{q}}\rangle\Big\},
\end{equation}
In the limit $j\rightarrow\infty$, we expect that the main
contribution to this integral arises from the neighborhood of the
identity for the group elements.  At the identity, the action reads
\begin{equation}
S_{f}^{\pm}[n,1]=2k_{f}\gamma^{\pm}\log\mbox{Tr}\Big[\prod_{q}^{\longrightarrow}\frac{1}{2}\big(1+\mathrm{i}\,n_{f,l_{q}}\cdot\sigma\big)\Big].
\end{equation}
This is the trace of a product of rank one projectors, its real part is always negative and vanishes when all the projectors are equal. Therefore, we expand the unit vectors $n_{f,l_{q}}$ around an unit vector $n_{f}$ common to all edges of the face,
\begin{equation}
n_{f,l_{q}}=n_{f}+\xi_{f,l_{q}}-\frac{(\xi_{f,l_{q}})^{2}}{2}n_{f}+O(\xi_{f,l_{q}})^{3},\qquad \mbox{with}\qquad n_{f}\cdot\xi_{f,l_{q}}=0,
\end{equation}
together with an expansion of the group elements around the identity
\begin{equation}
h_{v,l}=1-\frac{(A_{v,l})^{2}}{2}+\mathrm{i}\,\sigma\cdot A_{v,l}+O(A_{v,l})^{3}.
\end{equation}
The expansion of the action to second order follows the same steps as section \ref{self}. It is convenient to introduce
\begin{equation}
D_{f.v_{q}}^{\pm}=\epsilon_{v_{q},l_{q}}\eta_{l_{q},f}A_{v_{q},l_{q}}^{\pm}+\epsilon_{v_{q},l_{q+1}}\eta_{l_{q+1},f}A_{v_{q},l_{q+1}}^{\pm}\label{defD}
\end{equation}
and
\begin{equation}
\Phi_{f,v_{q}}^{\pm}=\eta_{l_{q},f}\eta_{l_{q+1},f}A_{v_{q},l_{q}}^{\pm}\wedge A_{v_{q},l_{q+1}}^{\pm}.\label{defPhi}
\end{equation}
After some algebra, the second order expansion of the action reads
\begin{align}
S_{f}^{\pm}[A_{v,l},\xi_{f,l}]&=k_{f}\gamma^{\pm}\sum_{q}\Big\{(n_{f}\wedge D_{f,l_{q}}^{\pm})^{2}+2\mathrm{i}\,n_{f}\cdot D_{f,v_{q}}^{\pm}+
2\mathrm{i}\,n_{f}\cdot \Phi_{f,v_{q}}^{\pm}\cr
&\qquad
-{\textstyle \frac{1}{2}}(\xi_{f,l_{q}})^{2}+{\textstyle \frac{1}{2}}\xi_{f,l_{q}}\cdot\xi_{f,l_{q+1}}+{\textstyle \frac{\mathrm{i}}{2}}\, n_{f}\cdot(\xi_{f,l_{q}}\wedge\xi_{f,l_{q+1}})\cr
&\qquad+\mathrm{i}\xi_{f,l_{q}}\cdot(D_{f,v_{q-1}}^{\pm}+D_{f,v_{q}}^{\pm})+\xi_{f,l_{q}}\cdot\big[n_{f}\wedge(D_{f,v_{q}}^{\pm}-D_{f,v_{q-1}}^{\pm})\big]
\Big\}.
\end{align}
In order to simplify the analysis, we perform a change of variable similar to \eqref{changeself},
\begin{equation}
A_{v,l}^{\pm}=A_{v,l}\pm\gamma^{\mp}X_{v,l}.
\end{equation}
Terms linear in $A_{v,l}^{\pm}$ are all of the form
\begin{equation}
\gamma^{+}L_{v,l}\cdot A_{v,l}^{+}+\gamma^{-}L_{v,l}\cdot A_{v,l}^{-}=L_{v,l}A_{v,l},\qquad\mbox{with }L_{v,l}\in{\Bbb R}^{3},
\end{equation}
so that they do not involve the variables $X_{v,l}$ Terms quadratic in $A_{v,l}^{\pm}$ are all of the form
\begin{equation}
\gamma^{+}B[A^{+}_{v,l},A^{+}_{v',l'}]+\gamma^{-}B[A^{-}_{v,l},A^{-}_{v',l'}],
\end{equation}
where the bilinear form $B[A^{\pm}_{v,l},A^{\pm}_{v',l'}]$ is either a scalar product $A^{\pm}_{v,l}\cdot A^{\pm}_{v',l'}$ or a wedge product
$n_{f}\cdot (A^{\pm}_{v,l}\wedge A^{\pm}_{v',l'})$. Substituting $A^{\pm}_{v,l}$ and $A^{\pm}_{v',l'}$, it is easily seen that
\begin{equation}
\gamma^{+}B[A^{+}_{v,l},A^{+}_{v',l'}]+\gamma^{-}B[A^{-}_{v,l},A^{-}_{v',l'}]=
B[A_{v,l},A_{v',l'}]+\gamma^{+}\gamma^{-}B[X_{v,l},X_{v',l'}].
\end{equation}
Then, we can express the total action $S_{f}=S_{f}^{+}+S_{f}^{-}$ as a sum of a BF type action
\begin{align}
S_{f}^{BF}[A,\xi]&=k_{f}\sum_{q}\Big\{(n_{f}\wedge D_{f,l_{q}})^{2}+2\mathrm{i}\,n_{f}\cdot D_{f,v_{q}}+
2\mathrm{i}\,n_{f}\cdot \Phi_{f,v_{q}}\cr
&\qquad\qquad
-{\textstyle \frac{1}{2}}(\xi_{f,l_{q}})^{2}+{\textstyle \frac{1}{2}}\xi_{f,l_{q}}\cdot\xi_{f,l_{q+1}}+{\textstyle \frac{\mathrm{i}}{2}}\, n_{f}\cdot(\xi_{f,l_{q}}\wedge\xi_{f,l_{q+1}})\cr
&\qquad\qquad+\mathrm{i}\xi_{f,l_{q}}\cdot(D_{f,v_{q-1}}+D_{f,v_{q}})+\xi_{f,l_{q}}\cdot\big[n_{f}\wedge(D_{f,v_{q}}-D_{f,v_{q-1}})\big]\Big\},
\end{align}
with $D^{\pm}_{f,v}$ and $\Phi_{f,v}$ as in \eqref{defD} and $\eqref{defPhi}$ but with $A_{v,l}$ instead of $A_{v,l}^{\pm}$, and an ultra local potential
\begin{equation}
Q_{f}[X]=\gamma^{+}\gamma^{-}k_{f}\sum_{q}\Big\{(n_{f}\wedge D_{f,l_{q}})^{2}+2\mathrm{i}\,n_{f}\cdot \Phi_{f,v_{q}}\Big\}.\label{defQ}
\end{equation}
To relate the BF face action to the more conventional one we encountered in \ref{BF}, let us perform the integration over the variables $\xi_{f.v_{q}}$, starting with $\xi_{f,v_{p}}$,
\begin{align}
&\int d\xi_{f,v_{p}}\exp jk_{f}\Big\{{\textstyle \frac{1}{2}}(\xi_{f,l_{p}})^{2}
+{\textstyle \frac{1}{2}}\xi_{f,l_{p}}\cdot(\xi_{f,l_{p-1}}+\xi_{f,l_{1}})+{\textstyle \frac{\mathrm{i}}{2}}\, \xi_{f,l_{p}}\cdot\big[ n_{f}\wedge(\xi_{f,l_{p-1}}-\xi_{f,l_{1}})\big]\cr
&\qquad\mathrm{i}\xi_{f,l_{p}}\cdot(D_{f,v_{p-1}}+D_{f,v_{p}})+\xi_{f,l_{p}}\cdot\big[n_{f}\wedge(D_{f,v_{p}}-D_{f,v_{p-1}})\big]
\Big\}
=\cr
&\frac{1}{j^{2}}\exp\frac{jk_{f}}{2}\Big\{{\textstyle \frac{1}{2}}(\xi_{f,l_{p-1}}+\xi_{f,l_{1}})+{\textstyle \frac{\mathrm{i}}{2}}\,\big[ n_{f}\wedge(\xi_{f,l_{p-1}}-\xi_{f,l_{1}})\big]\cr
&\qquad+\mathrm{i}\big[D_{f,v_{p-1}}+D_{f,v_{p}}-n_{f}\big(n_{f}\cdot(D_{f,v_{p-1}}+D_{f,v_{p}})\big)\big]+n_{f}\wedge(D_{f,v_{p}}-D_{f,v_{p-1}})
\Big\}^{2}=\cr
&\frac{1}{j^{2}}\exp jk_{f}\Big\{{\textstyle \frac{1}{2}}\xi_{f,l_{p-1}}\cdot\xi_{f,l_{1}}+{\textstyle \frac{\mathrm{i}}{2}}\,n_{f}\cdot(\xi_{f,l_{p-1}}\wedge\xi_{f,l_{1}})\cr
&\qquad\qquad-2 \mathrm{i}n_{f}\cdot(D_{f,v_{p-1}}\wedge D_{f,v_{1}})
-2\big(n_{f}\wedge D_{f,v_{p-1}})\,(n_{f}\wedge D_{f,v_{p-1}})\cr
&\qquad\qquad+\xi_{f,l_{p-1}}\cdot( \mathrm{i}D_{f,v_{p}}+n_{f}\wedge D_{f,v_{p}})
+\xi_{f,l_{1}}\cdot( \mathrm{i}D_{f,v_{p-1}}+n_{f}\wedge D_{f,v_{p-1}})\Big\}.
\end{align}
Note that  $\xi_{f,v_{p}}$ is orthogonal to $n_{f}$ so that it couples only to the projection of $D_{f,v_{p-1}}+D_{f,v_{p}}$ onto the subspace orthogonal to $n_{f}$. Gathering all the terms in the action pertaining to the edges $l_{q-1}$ and $l_{1}$, we get
\begin{align}
&-{\textstyle \frac{1}{2}}(\xi_{f,l_{q-1}})^{2}-{\textstyle \frac{1}{2}}(\xi_{f,l_{1}})^{2}+{\textstyle \frac{1}{2}}\xi_{f,l_{q-1}}\cdot\xi_{f,l_{1}}+{\textstyle \frac{\mathrm{i}}{2}}\, n_{f}\cdot(\xi_{f,l_{q-1}}\wedge\xi_{f,l_{1}})\cr
&\mathrm{i}\xi_{f,l_{p-1}}\cdot(D_{f,v_{p-2}}+D_{f,v_{p-1}}+D_{f,v_{p}})
+\xi_{f,l_{p-1}}\cdot\big[n_{f}\wedge(D_{f,v_{p-1}}-D_{f,v_{p-2}}+D_{f,v_{p}})\big]\cr
&\mathrm{i}\xi_{f,l_{1}}\cdot(D_{f,v_{p}}+D_{f,v_{1}}+D_{f,v_{p-1}})
+\xi_{f,l_{1}}\cdot\big[n_{f}\wedge(D_{f,v_{1}}-D_{f,v_{p-1}}-D_{f,v_{p}})\big]\cr
&-(D_{f,v_{p}}\wedge n_{f})^{2}-(D_{f,v_{p-1}}\wedge n_{f})^{2}-2(D_{f,v_{p}}\wedge n_{f})\cdot(D_{f,v_{p-1}}\wedge n_{f})\cr
&+2\mathrm{i}\,n_{f}\cdot(\Phi_{v_{q}}-D_{f,v_{p-1}}\wedge D_{f,v_{1}}).
\end{align}
The integration over the variable $\xi_{f,l_{p}}$ has a simple graphical interpretation. We have contracted the line $l_{q}$ and merged the vertex $v_{q}$ (associated with $D_{f,V_{q}}$) with the vertex $v_{q-1}$ (associated with $D_{f,v_{q-1}}$) into a new vertex (still called $v_{q-1}$), associated with $D_{f,v_{q-1}}+D_{f,v_{q}}$ and $\Phi_{f,v_{q}}-D_{f,v_{p-1}}\wedge D_{f,v_{1}}$. Therefore, we may pursue this procedure till we obtain a face with only two edges. Then, we proceed as in section \ref{self} for the self energy and integrate over $\xi_{f,1}-\xi_{f,2}$. The remaining variable $\xi_{f,1}+\xi_{f,2}$ is a Lagrange multiplier for the constraint $\sum_{q}D_{f,v_{q}}=0$, or explicitly
\begin{equation}
\sum_{1\leq q\leq p}\epsilon_{v_{q},l_{q}}\eta_{l_{q},f}A_{v_{q},l_{q}}+\epsilon_{v_{q},l_{q+1}}\eta_{l_{q+1},f}A_{v_{q},l_{q+1}}=0,
\end{equation}
which is nothing but the constraint \eqref{constraintdegenerate} written in terms of the variables $\epsilon_{v_{q-1},l_{q}}A_{v_{q-1},l_{q}}+\epsilon_{v_{q},l_{q}}A_{v_{q},l_{q}}$. Then, the rest of the discussion follows the same path as in section \ref{BF}: There are $2r$ independent constraints in the maximally degenerate case. The real part of the action vanishes identically as well as the imaginary part for a simply connected graph, once we have used these constraints and the closure constraints. Accordingly, we have
\begin{equation}
\int\, \prod_{v,l} dA_{v,l}\,\prod_{f,l}dn_{f,l}\,\exp j \Big\{\sum_{f}S_{f}^{BF}[A,n]\Big\}\quad\lesssim\quad j^{(\sum_{f}L_{f}-1)}\times j^{-2r},
\end{equation}
with $L_{f}$ the number of edges in the face $f$.

Let us finally analyze the ultra local terms given by the quadratic form $Q_{f}$ defined in \eqref{defQ}. Gathering the contributions of all faces, we get
\begin{equation}
Q[X]=\sum_{v}\Big\{\sum_{f}\gamma^{+}\gamma^{-}k_{f}
\big[n_{f}\wedge (A_{v,{\mathop{l}\limits^{\leftarrow}}_{\!f,v}}-A_{v,{\mathop{l}\limits^{\rightarrow}}_{\!f,v}})
\big]^{2}+2\mathrm{i}\,n_{f}\cdot (A_{v,{\mathop{l}\limits^{\leftarrow}}_{\!f,v}}\wedge A_{v,{\mathop{l}\limits^{\rightarrow}}_{\!f,v}})
\Big\},
\end{equation}
with ${\mathop{l}\limits^{\leftarrow}}_{\!f,v}$ (resp.${\mathop{l}\limits^{\rightarrow}}_{\!f,v}$) the edge entering (resp. leaving) the vertex $v$ along the face $f$. First we notice that this is a sum over all vertices of quadratic forms defined at each vertex involving only variables attached to that vertex. This is the reason why we called such a term "ultra local".

We then proceed as we did for the self-energy. The quadratic form has a real and an imaginary part, but its arguments are real. Therefore, the kernel of the associated bilinear form is the intersection of the kernel of the real part and of the imaginary part. Because the real part is a sum of squares, at each vertex and for each face we have
\begin{equation}
A_{v,{\mathop{l}\limits^{\leftarrow}}_{\!f,v}}=A_{v,{\mathop{l}\limits^{\rightarrow}}_{\!f,v}}\qquad\mbox{in the direction orthogonal to }n_{f}.
\end{equation}
Since in the maximally degenerate case all the $n_{f}$ are proportional to $n_{0}$, this simply implies that all vectors $A_{v,l}=C_{v}$  in the plane orthogonal to $n_{0}$, while the components collinear to $n_{0}$ are left unconstrained. Then, as in section \ref{self}, the closure constraints imply that $A_{v,l}=C_{v}$ also lies in the kernel of the  imaginary part. If we denote by $d_{v}$ the valence of vertex $v$ ($d_{v}$ may be lower than 5 since the external faces carrying spin 0 have to be removed), we get a rank of $2d_{v}-2$, (there are $3d_{v}$ variables and $2+d_{v}$ solutions), so that
\begin{equation}
\int \prod_{}dX\,\exp jQ(X)\quad\lesssim\quad j^{-\frac{\sum_{v}(2d_{v}-2)}{2}}.
\end{equation}
Taking all the terms together, we get
\begin{align}
A_{\cG}\quad
\lesssim\, \Lambda^{3F-3r +F+V -\sum_{v}d_{v}}.\label{maxdiv}
\end{align}
The first term is the power counting of the graph $\cG$ in BF theory with group $SU(2)$, while the second one results from a difference of normalization between EPRL/FK and BF theories. The last one is minus the half of sum of the ranks of the ultra local quadratic forms at each vertex. Since the latter are less important for non-degenerate configurations, we expect the maximally degenerate configuration to give a larger contribution, as long as the external spin remain finite. In particular, for the self-energy we have $d_{v}=4$ so that the maximally degenerate configurations are bounded by $\Lambda^{9}$. 

When applied to the self energy with $d_{v}=4$ because we set the external spins to 0, we get a bound in $\Lambda^{9}$, which suggests that degenerate configurations dominate in the EPRL model. However, this is only an upper bound since the zero modes of the degenerate configurations are not all gauge degrees of freedom, in particular the component of $A$ and $B$ along $n_{0}$ do not contribute to the action in the quadratic approximation. These modes require a more thorough study involving higher order terms. Nevertheless, using the asymptotic behavior of $6j$ symbols and fusion coefficients (see \cite{PRS}), we show in appendix \ref{degeneratespin} that degenerate configurations indeed dominate this correction to the self energy, but with an asymptotic behavior in $\Lambda^{7}$ instead of $\Lambda^{9}$. This is not in contradiction with the results of \cite{PRS}, since the latter use the relation $(6j)^{2}\sim\frac{1}{V}$,
 which implicitly assumes that the configuration is non-degenerate. Therefore, in the sum over spins we have to identify
 a  partial sum made of spins obeying a relation such that a maximally degenerate configurations exist. This partial sum
  behaves like $\Lambda^{7}$, while the remaining terms containing the non-degenerate configurations are in $\Lambda^{6}$.

\section{Concluding remark: hint on a phase transition}
\renewcommand{\theequation}{\thesection.\arabic{equation}}
\setcounter{equation}{0}

By parity, the $\phi^5$ 4-dimensional GFT has no two point function contribution to first order in the coupling constant.
At second order beyond the self-energy graph $\cG_2$, the only other graphs have tadpoles, hence
they are absent in the colored model; in the non-colored model they have less faces, so we can
expect the amplitude of $\cG_2$
to provide the dominant correction to the effective propagator
of the model.

Since that amplitude ${\cal A}_{\cG_2}$ is positive, we can expect the whole
self-energy correction $\Sigma$ to be also positive. The corresponding geometric
power series for the dressed or effective propagator
\be  C_{dressed} = C + C\Sigma C + C\Sigma C \Sigma C + ... = C (\frac {1}{1 - \Sigma C})
\ee
should therefore be singular when the spectrum of $\Sigma C $ has eigenvalue 1. This should
occur for $\lambda$ large enough, depending on the ultraviolet cutoff  $\Lambda$.
This is usually the signal of a phase transition. For instance in an ordinary
$\phi^4$ model a positive mass term corresponds to a double well potential
which signals a breaking of the $\phi \to -\phi$ symmetry. In the vector
$\phi^4$ ``Ginzburg-Landau'' model, it leads to the famous continuous
symmetry breaking with appearance of an associated Goldstone boson.

At a more speculative level, this hint of a phase transition support a scenario in
which ordinary macroscopic smooth space-time is an emergent phenomenon.
Group field theory, in particular its perturbative phase,
might be a more fundamental description and space-time might result from
condensation through a phase transition. This scenario is a version of what has
been called geometrogenesis. 

In this scenario the relationship of group field theory to space-time, gravitons and
general relativity would be somewhat similar to that between QCD
and effective theories of nuclear forces.

\section*{Acknowledgments}
T. K. and P. V. thank CPHT \'Ecole Polytechnique and LPT Orsay for
the hospitality. A. T. acknowledges the CNCSIS grant ``Idei''
454/2009, ID- 44 and the grant PN 09 37 01 02. P. V. acknowledges
the European Science Foundation Exchange Grant 2595 under the
Research Networking Program ``Quantum Geometry and Quantum Gravity''
and thanks the  Universit\`a di Napoli for partial financial suppport. We also acknowledge J. Ben Geloun, R. Gurau and S. Speziale for fruitful discussions.

\appendix

\section {Harmonic analysis on SU(2) and coherent states}

We include in this appendix well known formulas for self-completeness. We start with
\be
\int d g {R^{(j)}}^m_k (g)= \delta_{j0} \delta_{m 0} \delta_{k 0}
\ee
\be
{R^{(\jmath)}}^m_k (g) {R^{(\tilde\jmath})}^{\tilde m}_{\tilde k}
(g)=\sum_{J=|\jmath-\tilde \jmath|}^{\jmath+\tilde \jmath}
(\jmath,m;\tilde\jmath,\tilde m|J, M) (J,
K|\jmath,k;\tilde\jmath,\tilde k) {R^{(J)}}^M_K(g),
 \label{rprod}
\ee
where
 ${R^{(j)}}^m_k (g)$ are unitary representations of $SU(2)$ and $
(J, K|\jmath,k;\tilde\jmath,\tilde k)$ are the Clebsh-Gordan
coefficients.
We use the  normalizations of \cite{rovel}.

We have
\bea
\int d g \, {{\overline{R}^{(j)}}_n} \,^m (g) {R^{(j') p}}_q
(g)&=&
\frac{1}{d_j}\delta(j,j') \delta^m_q \delta_n^p \label{intgg}\\
\int d g \, {R^{(j) m}}_n (g) {R^{(j')p}}_q (g)&=&
\frac{1}{d_j}\delta(j,j') \epsilon^{m p}
\epsilon_{nq}\label{intgg-1}
\eea
with ${{\overline{R}^{(j)}}_n} \,^m (g)={R^{(j)m}}_n(g^{-1})$ which
imply
\bea
\int d g \tr_j Ag\,\tr_{j'}
Bg^{-1}&=&\frac{1}{d_j}\delta(j,j')\tr_j
AB
\label{gg-1}\\
\int d g \tr_j Ag\,\tr_{j'} Bg&=&\frac{1}{d_j}\delta(j,j')\tr_j
A\epsilon B^T\epsilon^T
\label{gg}
\eea
with $\epsilon\in SU(2)$,
\be \label{epsilonop}
\epsilon= \left( \begin{array}{cc}
0 & 1 \\
 -1 & 0 \\
\end{array}
 \right).
\ee
We have $\epsilon^T\epsilon = 1$ and $
\epsilon g \epsilon^T= \bar g
$.

\subsection{nj symbols} We have the {3j
symbols}
\be
\imath^{m_1 m_2 m_3}= \frac{(-1)^{j_1-j_2+m_3}}{\sqrt{d_{j_3}}}
(j_3, -m_3|j_1,m_1;j_2,m_2), \label{3j}
\ee
the { 6j symbols}
\be
\left(
                        \begin{array}{ccc}
                          j_1 & j_2 & j_3 \\
                          j_4 & j_5 & j_6 \\
                        \end{array}
                      \right) = \sum_{m_1.. m_6} \imath^{m_4 m_3 m_5} \imath_{m_5
                      m_1 m_6}
{\imath_{m_3}}^{ m_2 m_1} {\imath_{m_2 m_4 }}^{ m_6}, \label{6j}
\ee
and the { 9j symbols}
\be
\left\{\begin{array}{ccc}
                                            j_1 & j_2 & j_3 \\
                                            j_4 & j_5 & j_6 \\
                                            j_7 & j_8 & j_9
                                          \end{array}\right\}=
                                          \sum_{m_1.. m_6} \imath^{m_1 m_2 m_3} \imath^{m_4 m_5 m_6
}\imath^{m_7 m_8 m_9}\imath_{m_1 m_4 m_7} \imath_{m_2 m_5
m_8}\imath_{m_3 m_6 m_9}. \label{9j}
\ee
For the { 15j symbols} see \cite{rovel}.
The indices are raised and lowered with the tensor $\epsilon$.

\subsection{Coherent states}

Let us first consider the  ${\rm SU}(2)$ case.
We introduce the following parametrization for coherent states in
the spin-1/2 fundamental representation
\be
|\frac{1}{2}, n>= e^{i \theta \hat m \cdot \frac{\sigma}{2}}
|\frac{1}{2}, \frac{1}{2}>
\ee
with \be\hat m=(\sin\phi,-\cos\phi,0) \ee and $\sigma_i$ the Pauli
matrices
\bea
\sigma_1=\left(
      \begin{array}{cc}
        0 & 1 \\
        1 & 0 \\
      \end{array}
    \right)         \;\;\;
\sigma_2=\left(
      \begin{array}{cc}
        0 & -i \\
        i & 0 \\
      \end{array}
    \right)               \;\;\;
  \sigma_3=\left(
      \begin{array}{cc}
        1 & 0 \\
        0 &-1 \\
      \end{array}
    \right)
\eea
 This represents a rotation $g_n$ of the vector $n_0=(0,0,1)$ of
an angle $\theta$ around the $\hat m$ direction:
\be \label{gnrot}
n_0\longrightarrow
n=(\sin\theta\cos\phi,\sin\theta\sin\phi,\cos\theta)
\ee
with $\theta\in(0,\pi),\phi\in(0,2\pi)$. The coherent state $|j, n>$
is obtained in the same way replacing the generators $i \sigma_i/2$
with the appropriate operators $J_i$ in the $2j+1$ dimensional
representation. With this parametrization the scalar product of
coherent states reads
\begin{equation}
\langle j,n|j,\tilde{n}\rangle=\bigl(\cos\frac{\theta}{2}
\cos\frac{\tilde\theta}{2}+
\sin\frac{\theta}{2}\sin\frac{\tilde\theta}{2}
e^{-i(\phi-\tilde\phi)}\bigr)^{2j}\label{jnjn'}
\end{equation}
which implies
\be
\label{aj2}
\langle j,n|j,\tilde{n}\rangle= (\langle
{\textstyle \frac{1}{2}},n|{\textstyle \frac{1}{2}},\tilde{n}\rangle)^{2j}.
\ee

In the representation space $V^j$ of dimension $\rd_j\equiv 2j+1$:
we have
\be\label{ident-m1}
{\mathbf 1}_j = \sum_m |j,m \ket \bra j,m|,
\ee
where $|j,m\ket, m\in[-j,j]$ is the usual orthonormal basis in $V^j$.
We have:
\be
 {R^{(j') m}}_{m'}(g)
\equiv \bra j,m| g|j,m'\ket.
\ee
Hence
\be
\delta^m_{\; m'} = \rd_j \int_{{\rm SU}(2)} dg \,  {R^{(j) m}}_j (g)
{\overline{R}^{(j)}} \,^j_{\; m'} (g).
\label{mm'}
\ee

\section{Self-energy: comparison with \cite{PRS} and normalization conventions}
\renewcommand{\theequation}{\thesection.\arabic{equation}}
\setcounter{equation}{0}

We return to the ``self-energy'' graph $\cG_2$ of Fig. \ref{2P-fig}.
We first rewrite the propagator in a slightly different way, using
the gauge invariance. We perform a $SU(2)$ gauge transformation. We
multiply the $u^{\pm}, v^{\pm}$ variables simultaneously by $SU(2)$
elements $h,\tilde h$ which are the same for the left and right
components
\be
u^{\pm}\rightarrow h^{-1} u^{\pm},\;\;\;v^{\pm}\rightarrow \tilde
h^{-1} v^{\pm}
\ee
and we integrate over $h,\tilde h$ so that \eqref{defC} becomes:
\bea
C(g,g')&= &\int d h \,d \tilde h \,d u \,d v \,\prod_f
\sum_{j_f}\alpha_f\beta_f \int d n_f \sum_{m,\tilde m, k, \tilde k}\nn\\
&& \left(g'_{f+}v^{-1}_+|j_+m\rangle\langle j_+\tilde m|\,u_+ g_{f+}
R^{(j_+)}{\,^m}_{j_+}(\tilde h n_f) \,R^{(j_+)}{\,^{j_+}}_{\tilde
m}((
h n_f)^\dag)\right)\nn\\
&&\otimes\left(g'_{f-}v^{-1}_-|j_-k\rangle\langle j_-\tilde k|\,u_-
g_{f-} R^{(j_-)}{\,^k}_{j_-}(\tilde h n_f)
\,R^{(j_-)}{\,^{j_-}}_{\tilde k}(( h n_f)^\dag)\right) .
\eea
Note that we have also used \eqref{def}. Considering the tensor
product of representations \eqn{rprod} we get
\bea
C(g, g')&= &\int d h \,d \tilde h \,d u \,d v \,\prod_f
\sum_{j_f}\alpha_f\beta_f \int d n_f \sum_{m,\tilde m, k, \tilde k}
 (j_+ +j_-, M|j_+,
m;j_-, k)\nn\\
&&(j_+, \tilde m;j_-, \tilde k|j_+ +j_-, \tilde M)
 R^{(j_++j_-)}{\,^M}_{j_++j_-}(\tilde h n_f)
R^{(j_++j_-)}{\,^{j_++j_-}}_{\tilde M}(( h n_f)^\dag)\nn\\
&& (g'_{f+}v^{-1}_+|j_+m\rangle\langle j_+\tilde m|\,u_+
g_{f+})\otimes (g'_{f-}v^{-1}_-|j_-k\rangle \langle j_-\tilde
k|\,u_- g_{f-}) .
\eea
The integration over $n_f$ through \eqn{intgg-1} finally yields
\be
C(g,g')=\int d H d u d v\prod_{f=1}^4 \sum_{j_f}\alpha_{j_f}\tr
\left(u g_f (g'_f)^{-1 } v^{-1} T^\gamma_{j_f}(H)\right)
\label{newprop}
\ee
with  $ H=\tilde h h^\dag$, and
\bea
T^\gamma_{j_f}(H)&=& {\beta_{j_f}}\sum_{\stackrel{m\tilde m}{k\tilde
k}}
|j_{f+}m_f\rangle \langle j_{f+}\tilde m_f|\otimes |j_{f-}k_f> \nonumber \\
&&<j_{f-}\tilde k_f| \imath_{m_f k_f -M_f} \imath^{\tilde m_f \tilde
k_f -\tilde M_f} {R^{(j_{f+}+j_{f-}) M_f}}_{\tilde M_f}
(H)\label{TH}
\eea
where we have used \eqn{3j}. Using this expression and the amplitude
expression, it is checked below that this corresponds to the
normalizations of \cite{PRS}, with $k=2$.

Note that this choice $k=2$ is {\it not} the one advocated in
\cite{BRR}.

To prove this statement  we rewrite  the graph amplitude for the
"self-energy"  inserting the new expression of the propagator
\eqref{newprop}. We can neglect the open faces, since the external
legs have vanishing spin. Hence, we get
\be
{\mathcal A}({\mathcal G}_2)=\int  d H\,d u\, d v\,\prod_{a<b}
\mathcal{A}_{ab}(u, v,H),
\label{selfampl}
\ee
with the face amplitude
\bea
\mathcal{A}_{ab}&=&\sum_{j_{a}, j_{b}}  {\alpha
 }_{j_a}{\alpha
 }_{j_b}\int
d g_{  ab} d \tilde g_{ ab}
 \tr_{j_a +\otimes j_a -} \bigl( u_a  g_{ab} \tilde
g_{ab}^{-1}v_a^{-1} T_{j_a}^\gamma(H_a) \bigr) \nn\\
&&
 \tr_{j_b +\otimes j_b -} \bigl( u_b  g_{ab} \tilde
g_{ab}^{-1}v_b^{-1} T_{j_b}^\gamma(H_b) \bigr).
\eea
The amplitude for this ``self-energy'' graph is written in  \cite{PRS} as
\be
{\mathcal A}({\mathcal G}_2)=\sum_{j_{ab}\,\imath_a}\prod_{a<b}
d(j_{ab})\left(6j (j^+_{ab}) 6j (j^-_{ab})\right)^2(\prod_a f_a)^2,
\label{rovelampl}
\ee
where in \cite{PRS}, $d(j_{ab})\simeq j_{ab}^k$ in the ``ultraspin'' regime. One further has
\be
 6j(j_{ab})= \left(
                        \begin{array}{ccc}
                          j_{12} & j_{13} & j_{14} \\
                          j_{23} & j_{24} & j_{34} \\
                        \end{array}
                      \right)
\ee
defined as in Eq. \eqn{6j}, while
\be
f_1=\sqrt{d_{j_{12}}d_{j_{13}}d_{j_{14}}} \left\{\begin{array}{ccc}
                                            j^+_{12 }& j^+_{14} & j^+_{13} \\
                                            j^-_{12 }& j^-_{14} & j^-_{13} \\
                                            j_{12 }& j_{14} & j_{13}
                                          \end{array}\right\}
\ee
and cyclically for $f_2, f_3,f_4$. Both expressions, \eqn{selfampl}
and \eqn{rovelampl} are calculated for external $j's$ put to zero,
that is, the contributions of faces with external legs are put to
$1$.

To compare our expression to \eqref{rovelampl},
we perform the integration on the variables $u,v$ and we
rewrite the integrand of face amplitudes in the form:
\be
  \left(  g^{\pm}_{ab} {\tilde{ g}^{\pm}_{ab}}\,^{-1}\right)^{m_{ab}}_{\, k_{ab}}  \left({v^{\pm}_a}^{-1}\right)^{k_{ab}}_{\; p_{ab}}
  \left(T_{j_a}^\gamma\right)^{p_{ab}}_{\, q_{ab}}(H_a) \left(u^{\pm}_a\right)^{q_{ab}}_{\, m_{ab}},
\ee
where we use the shorthand notation
\bea
(g)^m_n=R^{(j)}{}^m_{\;\; n}(g).
\eea
We need to perform  integrals of the form
\be
\int d u_a^{\pm} \prod_{b\ne a}\left(u^{\pm}_a\right)^{q_{ab}}_{\,
m_{ab}}
\ee
\be
\int d v_a^{\pm} \prod_{b\ne
a}\left({v^{\pm}_a}^{-1}\right)^{k_{ab}}_{\, p_{ab}}
\ee
with $a,b=1,...,4$ (we have 16 integrals in total). Using
\eqn{rprod} and \eqn{intgg} we obtain for $a=1$
\be
\int d u_1^{\pm} \left(u^{\pm}_1\right)^{q_{12}}_{\, m_{12}}
\left(u^{\pm}_1\right)^{q_{13}}_{\, m_{13}}
\left(u^{\pm}_1\right)^{q_{14}}_{\, m_{14}}=
\imath_{\pm}^{q_{12}q_{13}q_{14}}\imath^{\pm}_{m_{12}m_{13}m_{14}}\label{uint}
\ee
\be
\int d v_1^{\pm} \left({v^{\pm}_1}^{-1}\right)^{k_{12}}_{\, p_{12}}
\left({v^{\pm}_1}^{-1}\right)^{k_{13}}_{\, p_{13}}
\left({v^{\pm}_1}^{-1}\right)^{k_{14}}_{\, p_{14}}=
\imath_{\pm}^{k_{12}k_{13}k_{14}}\imath^{\pm}_{p_{12}p_{13}p_{14}}\label{vint}
\ee
and similar results for $a=2,3,4$. All indices are double for $+$
and $-$ variables, that is, they should carry an extra superscript
(e.g. $m_{ab}\rightarrow m^{\pm}_{ab}$). We now perform the
integration on the variables $g, \tilde g$. For each face ${\mathcal
A}_{ab}$ they appear twice, once attached to the propagator
containing the $a$ variables, once to the propagator containing the
$b$ ones. This explains the switch in the indices below. We have 6
integrals to perform for each $SU(2)$ copy. By means of \eqn{intgg}
we find
\be
\int d g_{ab}^{\pm}d {\tilde g}_{ab}^\pm \left(g_{ab}^{\pm}
{\tilde g}_{ab}^{-1}\right)^{m_{ab}}_{k_{ab}} \left(g_{ab}^{\pm}
{\tilde g}_{ab}^{-1}\right)^{m_{ba}}_{k_{ba}} =
\frac{1}{d_{j^+_{ab}} d_{j^-_{ab}}} \epsilon^{m_{ab} m_{ba}}
\epsilon_{k_{ab} k_{ba}} . \label{intggtilde}
\ee
To compare to the expression of \cite{PRS} we still have to perform
the integration over $H_a$ appearing in the $T^\gamma_j$. From
\eqn{TH} we have
\be
(T^\gamma_{j_{a}})^{p_{ab}}_{q_{ab}}(H_a)=  \beta_a
{\imath^{p_{ab}^+ p_{ab}^-}}_{M _{ab}} {\imath_{q_{ab}^+
q_{ab}^-}}^{ \tilde M_{ab}} (H_{a})^{M_{ab}}_{\tilde M_{ab}}.
\label{tja}
\ee
Therefore we have 4 integrals of the form
\be
\beta_a\prod_{b\ne a}\left({\imath^{p_{ab}^+ p_{ab}^-}}_{M _{ab}}
{\imath_{q_{ab}^+ q_{ab}^-}}^{ \tilde M_{ab}} \right)  \int d
H_a\,\prod_{b\ne a}(H_{a})^{M_{ab}}_{\tilde M_{ab}}. \label{intHa}
\ee
We obtain
\be\int d
H_1\,(H_{1})^{M_{12}}_{\tilde M_{12}}\,(H_{1})^{M_{13}}_{\tilde
M_{13}}(H_{1})^{M_{14}}_{\tilde M_{14}}=\imath^{M_{12}M_{13}M_{14}}
\imath_{\tilde M_{12}\tilde M_{13}\tilde M_{14}} \label{intH}
\ee
and similarly for the others. We replace this result into
\eqn{intHa} for each $a\in\{1,..,4\}$. We replace then all
integration results \eqn{uint},
\eqn{vint}\eqn{intggtilde} \eqn{intHa} into the
expression for the graph amplitude \eqn{selfampl} and we obtain
\be
{\mathcal A}({\mathcal G}_2)=
\left(\prod_{a<b}
d_{j^+_{ab}}d_{j^-_{ab}}\right)\; \left(6j_{ab}^+
\;6j_{ab}^-\right)^2 \left (\prod_{a=1}^4  f_a\right)^2 .\label{append}
\ee
As already stated above, this reproduces \eqn{rovelampl} with $k=2$.

\section{Asymptotics of $6j$ and\\
 fusion coefficients in the degenerate case}
 
 \label{degeneratespin}

In this appendix, we investigate the asymptotic behavior of the self-energy correction using sum over spins. We first derive the asymptotics of the $6j$ symbols which yields the power counting in the $SU(2)$ BF theory and then fusion coeffcients $f$ appearing in \eqref{append} to obtain the power counting in the EPRL model. 

\subsection{Degenerate $6j$ and BF theory}

In the general case, the $6j$ symbols can be written using Racah's single sum formula (see for instance \cite{Razvan6j})
\begin{equation}
\begin{Bmatrix}
j_{12}&j_{23}&j_{13}\cr
j_{34}&j_{14}&j_{24}\end{Bmatrix}
=
\Delta(1,2,3)\Delta(1,2,4)\Delta(1,3,4)\Delta(2,3,4)\times\sum_{k}\frac{(-1)^{k}(k+1)!}{F(k)}
\end{equation}
with
\begin{equation}
\Delta(a,b,c)=\Bigg(
\frac{(-j_{ab}+j_{bc}+j_{ac})!(j_{ab}-j_{bc}+j_{ac})!(j_{ab}+j_{bc}-j_{ac})!}{(j_{ab}+j_{bc}+j_{ac}+1)!}
\Bigg)^{\frac{1}{2}}
\end{equation}
and
\begin{multline}
 F(k)=(k\!-\!j_{12}\!-\!j_{23}\!-\!j_{13})!(k\!-\!j_{13}\!-\!j_{34}\!-\!j_{14})!(k\!-\!j_{23}\!-\!j_{34}\!-\!j_{24})!(k\!-\!j_{12}\!-\!j_{24}\!-\!j_{14})!\cr
\times(j_{12}\!+\!j_{23}\!+\!j_{34}\!+\!j_{14}\!-\!k)!(j_{12}\!+\!j_{13}\!+\!j_{34}\!+\!j_{24}\!-\!k)!(j_{23}\!+\!j_{13}\!+\!j_{24}\!+\!j_{14}\!-\!k)!
\end{multline}
The sums runs over all integers $k$ such that the arguments of the factorials are non negative.
 
Consider a degenerate tetrahedron which is reduced to a single edge, whose vertices are labelled 1,2,3,4 with 1 and 4 on the boundary of the edge. The associated spin (lengths of the edges of the tetrahedron) between vertices $a$ and $b$ ($a<b$) is $j_{ab}$ and we have $j_{ac}=j_{ab}+j_{bc}$ if $a<b<c$. Therefore only the three spins $j_{12}, j_{23},j_{34}$ are independent and we have
\begin{multline}
 F(k)=(k-2j_{12}-2j_{23})!(k-2j_{12}-2j_{23}-2j_{34})!(k-2j_{23}-2j_{34})!(k-2j_{12}-2j_{23}-2j_{34})!\cr
\times (2j_{12}+2j_{23}+2j_{34}-k)!(2j_{12}+2j_{23}+2j_{34}-k)!(2j_{12}+4j_{23}+2j_{34}-k)!
\end{multline}
The sum over $k$ is restricted to the single term $k=2(j_{12}+j_{23}+j_{34})$
and we have
\begin{equation}
 F(k)=(2j_{12})!(2j_{34})!(2j_{23})!.
\end{equation}
There are also simplifications in the factors $\Delta$,
\begin{multline}
\Delta(1,2,3)=\Bigg(
\frac{(2j_{23})!(2j_{12})!}{(2j_{12}+2j_{23}+1)!}
\Bigg)^{\frac{1}{2}},
\Delta(1,2,4)=\Bigg(
\frac{(2j_{12})!(2j_{23}+2j_{34})!}{(2j_{12}+2j_{23}+2j_{34}+1)!}\Bigg)^{\frac{1}{2}},
\cr
\Delta(2,3,4)=\Bigg(
\frac{(2j_{34})!(2j_{23})!}{(2j_{34}+2j_{23}+1)!}
\Bigg)^{\frac{1}{2}},
\Delta(1,3,4)=\Bigg(
\frac{(2j_{34})!(2j_{12}+2j_{23})!}{(2j_{12}+2j_{23}+2j_{34}+1)!}\Bigg)^{\frac{1}{2}}.
\end{multline}
Taking all the terms together we get
\begin{multline}
\begin{Bmatrix}
j_{12}&j_{23}&j_{13}\cr
j_{34}&j_{14}&j_{24}\end{Bmatrix}
=(-1)^{2(j_{12}+j_{23}+j_{34})}\cr
\times\Bigg(
\frac{(2j_{23})!(2j_{12})!}{(2j_{12}+2j_{23}+1)!}
\frac{(2j_{34})!(2j_{23})!}{(2j_{34}+2j_{23}+1)!}
\frac{(2j_{34})!(2j_{12}+2j_{23})!}{(2j_{12}+2j_{23}+2j_{34}+1)!}
\frac{(2j_{12})!(2j_{23}+2j_{34})!}{(2j_{12}+2j_{23}+2j_{34}+1)!}
\Bigg)^{\frac{1}{2}}\cr
\times\frac{(2j_{12}+2j_{23}+2j_{34}+1)!}
{(2j_{12})!(2j_{23})! (2j_{34})!}
\end{multline}
which simplifies into
\begin{equation}
\begin{Bmatrix}
j_{12}&j_{23}&j_{13}\cr
j_{34}&j_{14}&j_{24}\end{Bmatrix}
=\frac{(-1)^{2(j_{12}+j_{23}+j_{34})}}{\sqrt{2j_{12}+2j_{23}+1}\sqrt{2j_{34}+2j_{23}+1}}.
\end{equation} 
This yields an asymptotic behavior ($k_{ab}\in[0,1]$ fixed)
\begin{equation}
\begin{Bmatrix}
jk_{12}& jk_{23}& jk_{13}\cr
jk_{34}&jk_{14}&jk_{24}\end{Bmatrix}^{2}
\mathop{\sim}_{j\rightarrow\infty}\frac{1}{j^{2}}.
\end{equation}
Consequently, the degree of divergence of the $SU(2)$ BF theory graph (double sunset without external legs) made of two vertices (each with one $6j$), three edges and 6 faces (each with one $dj=2j+1$) is
\begin{equation}
d^{\mbox{\tiny degenerate}}_{\mbox{\tiny BF}}=3+6-2=7<d^{\mbox{\tiny non degenerate}}_{\mbox{\tiny BF}}=9,
\end{equation} 
where we sum over only three spins in the maximally degenerate case. Let us note that this is less than the degree of divergence of non degenerate configurations, so that the latter are dominant in BF theory, at least for this graph.
 
 \subsection{Degenerate fusion coefficients and the EPRL model}
 
 Using the notations of \cite{PRS} (Appendix B), the fusion coefficients can be expressed as a product of a $9j$ and a $3j$ coefficient,
 
\begin{multline}f^{i^{+}i^{-}}_{i}(j_{1},j_{2},j_{3},0)\quad=\quad\delta_{i^{+},j_{3}^{+}}\delta_{i^{-},j_{3}^{-}}\delta_{i,j_{3}}\sqrt{d_{j_{1}}d_{j_{2}}d_{j_{3}}}\times\cr
\hskip-0.7cm\Bigg(\frac{(2j_{1}^{+})!(2j_{1}^{-})!(2j_{2}^{+})!(2j_{2}^{-})!(j_{1}^{+}\!+\!j_{1}^{-}\!+\!j_{2}^{+}\!+\!j_{2}^{-}\!-\!j_{3})!(j_{1}^{+}\!+\!j_{1}^{-}\!+\!j_{2}^{+}\!+\!j_{2}^{-}\!+\!j_{3}\!+\!1)!}
{(2j_{1}^{+}\!+\!2j_{1}^{-}\!+\!1)!(2j_{2}^{+}\!+\!2j_{2}^{-}\!+\!1)!
(j_{1}^{+}\!+\!j_{2}^{+}\!-\!j_{3}^{+})!(j_{1}^{+}\!+\!j_{2}^{+}\!+\!j_{3}^{+}\!+\!1)!(j_{1}^{-}\!+\!j_{2}^{-}\!-\!j_{3}^{-})!(j_{1}^{-}\!+\!j_{2}^{-}\!+\!j_{3}^{-}\!+\!1)!}\Bigg)^{\frac{1}{2}}
\cr
\times
\Bigg(\frac{(2j_{3}^{+})!(2j_{3}^{-})!(j_{3}^{+}\!+\!j_{3}^{-}\!-\!j_{1}^{+}\!-\!j_{1}^{-}\!+\!j_{2}^{+}\!+\!j_{2}^{-})!(j_{3}^{+}\!+\!j_{3}^{-}\!+\!j_{1}^{+}\!+\!j_{1}^{-}\!-\!j_{2}^{+}\!-\!j_{2}^{-})!}
{(1\!+\!2j_{3}^{+} \!+\!2j_{3}^{-})!(j_{3}^{+}\!-\!j_{1}^{+}\!+\!j_{2}^{+})!(j_{3}^{+}\!+\!j_{1}^{+}\!-\!j_{2}^{+})!(j_{3}^{-}\!-\!j_{1}^{-}\!+\!j_{2}^{-})!(j_{3}^{-}\!+\!j_{1}^{-}\!-\!j_{2}^{-})!}
\Bigg)^{\frac{1}{2}},
\end{multline}
with as usual $j^{\pm}_{a}=\frac{1\pm\gamma}{2}j_{a}$. The second factor is the contribution from the $9j$ while the third one is that of the $3j$. Let us notice that if $\gamma=1$, then $j_{a}^{+}=j_{a}$ and $j_{a}^{-}=0$, so that $f^{i^{+}i^{-}}_{i}(j_{1},j_{2},j_{3},0)\quad=\quad\delta_{i^{+},j_{3}^{+}}\delta_{i^{-},j_{3}^{-}}\delta_{i,j_{3}}$. Thus, the theory reduces to a $SU(2)$ BF theory using \eqref{append}.

The result is symmetrical in the indices $1,2,3$, so let us write the degeneracy condition on the triangle as $j_{1}+j_{2}=j_{3}$ to eliminate $j_{3}$. After some simplifications, we get 
\begin{equation}
f^{i^{+}i^{-}}_{i}(j_{1},j_{2},j_{3},0)\quad=\quad\delta_{i^{+},j_{3}^{+}}\delta_{i^{-},j_{3}^{-}}\delta_{i,j_{3}}\sqrt{\frac{2j_{1}+2j_{2}+1}{(2j_{1}^{+}+2j_{2}^{+}+1)(2j_{1}^{-}+2j_{2}^{-}+1)}}.
\end{equation}
Thus, the fusion coefficients scale as
\begin{equation}
f^{i^{+}i^{-}}_{i}(jk_{1},jk_{2},jk_{3})\mathop{\sim}_{j\rightarrow\infty}\frac{\delta_{jk^{+},jk_{3}^{+}}\delta_{i^{-},jk_{3}^{-}}\delta_{i,j_{3}}}{\sqrt{j}}.
\end{equation}
Accordingly, the power counting of the maximally degenerate configurations is (the summation over the intertwiners $i$ is trivial thanks to the Kronecker symbols)
\begin{equation}
d^{\mbox{\tiny degenerate}}_{\mbox{\tiny EPRL}}=3+12+2\Big(2\times(-1)+4\times\frac{-1}{2}\Big)=7>d^{\mbox{\tiny non degenerate}}_{\mbox{\tiny EPRL}}=6,
\end{equation} 
the first term is the contribution of the $6j$ (two per vertices) and the second one the contribution from the fusion coefficients (4 per vertices).
Therefore, the degenerate configurations dominate in the EPRL model for this graph, in accordance with the quadratic approximation. Nevertheless, there is no reason to believe that this is a general feature of the model, since the quadratic approximation only yields an upper bound.


\begin{thebibliography}{99}


\bibitem{boul}
D.~V.~Boulatov, Mod. Phys. Lett. {\bf  A7} (1992) 1629;
eprint {\tt hep-th/9202}
H.~Ooguri, Mod. Phys. Lett. {\bf  A7} (1992) 2799;
eprint {\tt hep-th/9205090}.

\bibitem{Freidel}
  L.~Freidel,
  ``Group field theory: An overview,''
  Int.\ J.\ Theor.\ Phys.\  {\bf 44}, 1769 (2005)
  [arXiv:hep-th/0505016].

\bibitem{oriti}
  D.~Oriti,
  ``The group field theory approach to quantum gravity,''
  arXiv:gr-qc/0607032.




\bibitem{rovel}
C.~Rovelli, {\it Quantum Gravity}
(Cambridge University Press, Cambridge, 2004).

\bibitem{FreiLoua}
  L.~Freidel and D.~Louapre,
  ``Non-perturbative summation over 3D discrete topologies,''
  Phys.\ Rev.\  D {\bf 68}, 104004 (2003)
  [arXiv:hep-th/0211026].

\bibitem{MNRS}
J.~Magnen, K.~Noui, V.~Rivasseau and M.~Smerlak,
``Scaling behaviour of three-dimensional group field theory,''
Class.\ Quant.\ Grav.\  {\bf 26}, 185012 (2009)
[arXiv:0906.5477 [hep-th]].

\bibitem{EPR}
  J.~Engle, R.~Pereira and C.~Rovelli,
  ``The loop-quantum-gravity vertex-amplitude,''
  Phys.\ Rev.\ Lett.\  {\bf 99}, 161301 (2007)
  [arXiv:0705.2388 [gr-qc]].

\bibitem{LS}
  E.~R.~Livine and S.~Speziale,
  ``A new spinfoam vertex for quantum gravity,''
  Phys.\ Rev.\  D {\bf 76}, 084028 (2007)
  [arXiv:0705.0674 [gr-qc]].

\bibitem{FreKra}
  L.~Freidel and K.~Krasnov,
  ``A New Spin Foam Model for 4d Gravity,''
  Class.\ Quant.\ Grav.\  {\bf 25}, 125018 (2008)
  [arXiv:0708.1595 [gr-qc]].

\bibitem{ELPR}
  J.~Engle, E.~Livine, R.~Pereira and C.~Rovelli,
  ``LQG vertex with finite Immirzi parameter,''
  Nucl.\ Phys.\  B {\bf 799}, 136 (2008)
  [arXiv:0711.0146 [gr-qc]].

\bibitem{PRS}
  C.~Perini, C.~Rovelli and S.~Speziale,
  ``Self-energy and vertex radiative corrections in LQG,''
  Phys.\ Lett.\  B {\bf 682}, 78 (2009)
  [arXiv:0810.1714 [gr-qc]].


\bibitem{linhom}
  J.~B.~Geloun, T.~Krajewski, J.~Magnen and V.~Rivasseau,
``Linearized Group Field Theory and Power Counting Theorems,''
  arXiv:1002.3592 [hep-th].

\bibitem{BonSme}
  V.~Bonzom and M.~Smerlak,
  ``Bubble divergences from cellular homology,''
  arXiv:1004.5196 [gr-qc].

\bibitem{fgo}
  L.~Freidel, R.~Gurau and D.~Oriti,
  ``Group field theory renormalization - the 3d case: power counting of
  divergences,''
  Phys.\ Rev.\  D {\bf 80}, 044007 (2009)
  [arXiv:0905.3772 [hep-th]].



\bibitem{FreidelConrady}
  F.~Conrady and L.~Freidel,
  Phys.\ Rev.\  D {\bf 78} (2008) 104023
  [arXiv:0809.2280 [gr-qc]].

\bibitem{gurau1}
  R.~Gurau,
``Colored Group Field Theory,''
  arXiv:0907.2582 [hep-th].
\bibitem{Tana1}
  A.~Tanasa,
  ``Algebraic structures in quantum gravity,''
  Class.\ Quant.\ Grav.\  {\bf 27}, 095008 (2010)
  [arXiv:0909.5631 [gr-qc]].
\bibitem{gurau2}
  R.~Gurau,
``Topological Graph Polynomials in Colored Group Field Theory,''
  arXiv:0911.1945 [hep-th].

\bibitem{gurau3}
  R.~Gurau,
  ``Lost in Translation: Topological Singularities in Group Field Theory,''
  arXiv:1006.0714 [hep-th].



\bibitem{Ben1}
J.~Ben~Geloun, J.~Magnen and V.~Rivasseau,
``Bosonic Colored Group Field Theory,''
arXiv:0911.1719 [hep-th].

\bibitem{perelomov}
  A.~M.~Perelomov,
  ``Generalized coherent states and their applications,''
{  Berlin, Springer (1986)}

\bibitem{gr}
R.~Gurau and V.~Rivasseau,
``Parametric representation of noncommutative field theory,''
Commun.\ Math.\ Phys.\  {\bf 272}, 811 (2007)
[arXiv:math-ph/0606030].

\bibitem{RT}
  V.~Rivasseau and A.~Tanasa,
  ``Parametric representation of `critical' noncommutative QFT models,''
  Commun.\ Math.\ Phys.\  {\bf 279} (2008) 355
  [arXiv:math-ph/0701034].


\bibitem{KRTW}
  T.~Krajewski, V.~Rivasseau, A.~Tanasa and Z.~Wang,
  ``Topological Graph Polynomials and Quantum Field Theory, Part I: Heat Kernel
  Theories,''
  arXiv:0811.0186 [math-ph].

\bibitem{KRV}
  T.~Krajewski, V.~Rivasseau and F.~Vignes-Tourneret,
 ``Topological graph polynomials and quantum field theory, Part II: Mehler
  kernel theories,''
  arXiv:0912.5438 [math-ph].

\bibitem{BRR}
  E.~Bianchi, D.~Regoli and C.~Rovelli,
  ``Face amplitude of spinfoam quantum gravity,''
  arXiv:1005.0764 [Unknown].

\bibitem{Baez}
  J.~C.~Baez,
  ``An introduction to spin foam models of BF theory and quantum gravity,''
  Lect.\ Notes Phys.\  {\bf 543}, 25 (2000)
  [arXiv:gr-qc/9905087].

\bibitem{Razvan6j}
R.~Gurau,
  "The Ponzano-Regge asymptotic of the 6j symbol: An Elementary proof'',
  Annales Henri Poincare {\bf 9 } (2008)  1413-1424
   arXiv:0808.3533 [math-ph].


\end{thebibliography}
\end{document}